\documentclass[prd,preprint,superscriptaddress]{revtex4}
\usepackage[dvips]{graphicx}
\usepackage{epsfig}
\usepackage{amssymb}
\usepackage{appendix}
\usepackage{amsmath}

\def\lsim{\mathrel{\hbox{\rlap{\hbox{\lower4pt\hbox{$\sim$}}}\hbox{$<$}}}}

\makeatletter

\newcommand{\Rmnum}[1]{\expandafter\@slowromancap\romannumeral #1@}
\makeatother

\begin{document}

\title{Daily modulation due  to channeling in direct dark matter crystalline detectors}

\author{Nassim Bozorgnia}
\email{nassim@physics.ucla.edu}
\affiliation{Department of Physics and Astronomy, UCLA, 475 Portola Plaza, Los
  Angeles, CA 90095, USA}

\author{Graciela B. Gelmini}
\email{gelmini@physics.ucla.edu}
\affiliation{Department of Physics and Astronomy, UCLA, 475 Portola Plaza, Los
  Angeles, CA 90095, USA}

\author{Paolo Gondolo}
\email{paolo@physics.utah.edu}
\affiliation{Department of Physics, University of Utah, 115 South 1400 East \# 201,
  Salt Lake City, UT 84112, USA}


\begin{abstract}

The channeling of the ion recoiling after a collision with a WIMP in direct dark matter crystalline detectors
 produces a larger scintillation or ionization signal than otherwise expected. Channeling is a directional effect which depends on
 the velocity distribution of WIMPs in the dark halo of our Galaxy and could lead to a daily modulation of the signal.
 Here we compute upper bounds to the expected amplitude of daily modulation due to channeling using channeling fractions that we obtained with
 analytic models in prior work. After developing the general formalism, we examine the possibility of finding a daily modulation due to channeling   in the data already collected by the DAMA/NaI and DAMA/LIBRA experiments.  We find that even the largest daily modulation amplitudes (of the order of 10\% in some instances) would not be observable for WIMPs in the standard halo in the 13 years of data taken by the DAMA collaboration. For these to be observable the DAMA total rate should be 1/40 of what it is or the total DAMA exposure should be 40 times larger.  The daily modulation due to channeling will be difficult to measure in future experiments. We find it could be observed for light WIMPs in solid Ne, assuming no background.

 \end{abstract}

\maketitle
\section{Introduction}

The channeling  effect in crystals refers to the orientation dependence of charged ion penetration in crystals. Channeling occurs when ions propagating in a crystal along symmetry axes and planes suffer a series of small-angle scatterings  that maintain them in the open ``channels"  in between the rows or planes of lattice atoms and thus penetrate much further into the crystal than in other directions and loose all their energy into electrons.
In dark matter crystalline detectors, a channeled ion recoiling after a collision with a WIMP (Weakly Interacting Massive Particle) would give all its energy to electrons, thus the quenching factor is $Q\simeq 1$ instead of the usual $Q< 1$ for a non-channeled ion. Thus channeling  increases the ionization or scintillation signal expected from a WIMP. The  potential importance of the channeling effect for direct  dark matter detection was first pointed out for NaI (Tl) by Drobyshevski~\cite{Drobyshevski:2007zj} and subsequently by the DAMA collaboration~\cite{Bernabei:2007hw} in 2007. In 2008, Avignone, Creswick, and Nussinov~\cite{Avignone:2008cw} suggested that a daily modulation due to channeling could occur in  NaI crystals,  which would be a background free dark matter signature. Such a  modulation of the rate due to channeling is expected to occur at some level because the ``WIMP wind" arrives to Earth  on average from a particular direction fixed to the Galaxy. Assuming that the dark matter halo is on average at rest with respect to the Galaxy, this is the direction towards  which the Earth  moves with respect to the Galaxy. Earth's daily rotation naturally changes the direction of the ``WIMP wind'' with respect to the crystal axes, thus changing the amount of recoiling ions that are channeled vs non-channeled. This  amounts to a daily modulation of the dark matter signal detectable via scintillation or ionization.

Using analytic models of channeling which started to be developed in the 1960's, shortly after the effect was discovered, mostly by Lindhard~\cite{Lindhard:1965} and collaborators,  we  recently computed channeling probabilities as function of the recoil energy $E_R$ and initial direction  $\hat{\bf q}$ of a recoiling ion in different materials~\cite{BGGI, BGG}. We used a recursion of the addition rule in probability theory (see Eq. 5.13 in Ref.~\cite{BGGI}) to find the probability $\chi(E_R,\hat{\bf q})$ that a recoiling ion enters into any channel in terms of the channeling fractions for single channels  $\chi_i(E_R,\hat{\bf q})$ that we computed (where the index i runs over all channels, both axial and planar).  The channeling fractions for axial and planar channels  are given in Eqs. 5.2 and 5.4 of Ref.~\cite{BGGI}, respectively.

In our previous papers~\cite{BGGI, BGG}, we also obtained the ``geometric" channeling fraction $P_{\rm geometric}(E_R)$ in the crystals we studied, by  averaging the channeling probability $\chi(E_R,\hat{\bf q})$ over the initial recoil directions $\hat{\bf q}$ (assuming an isotropic distribution in $\hat{\bf q}$)
\begin{equation}
P_{\rm geometric}(E_R)=\frac{1}{4\pi}\int{\chi(E_R, \hat{\bf q})d\Omega_q}.
\label{geometric-fraction}
\end{equation}
This integral was computed using the  Hierarchical Equal Area iso-Latitude Pixelization (HEALPix) ~\cite{HEALPix:2005} of the recoil direction sphere (see Appendix B of Ref.~\cite{BGGI}). Here ``geometric'' refers to assuming that the distribution of recoil directions is isotropic. In reality, in a dark matter direct detection experiment, the distribution of recoil directions depends on the momentum distribution of the incoming WIMPs (see Section II).

\begin{figure}
\begin{center}
  \includegraphics[height=137pt]{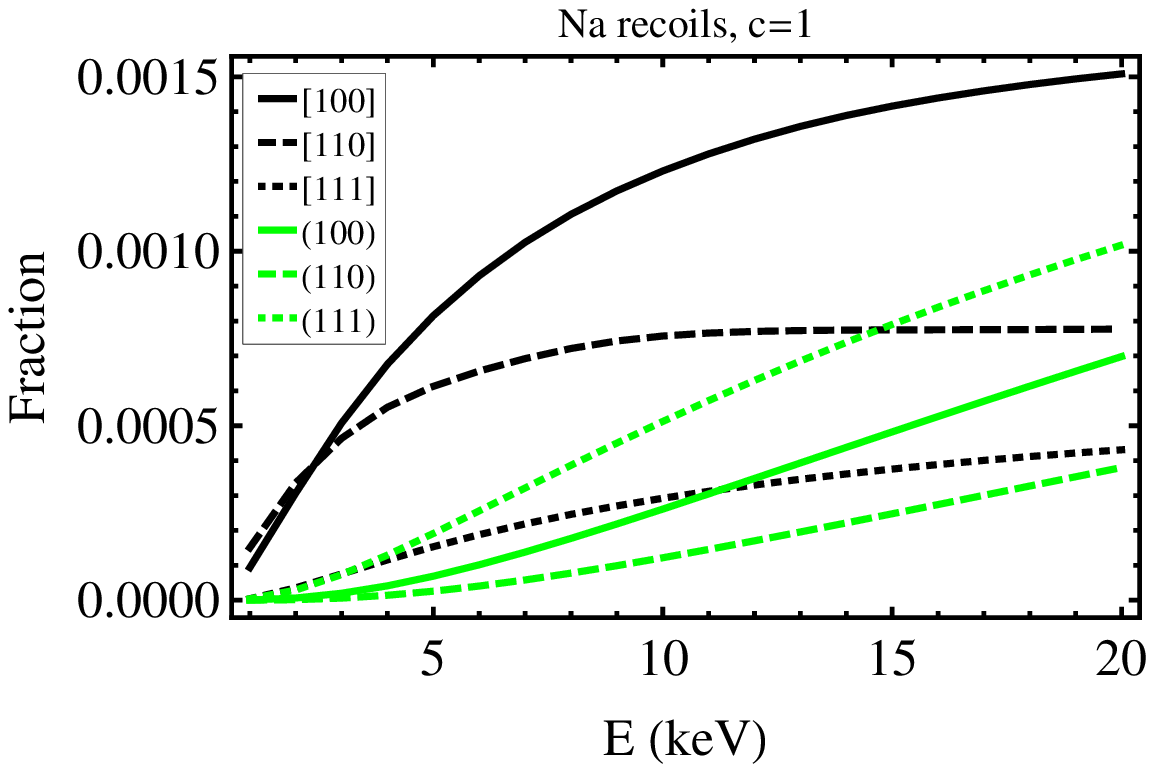}
  \includegraphics[height=144pt]{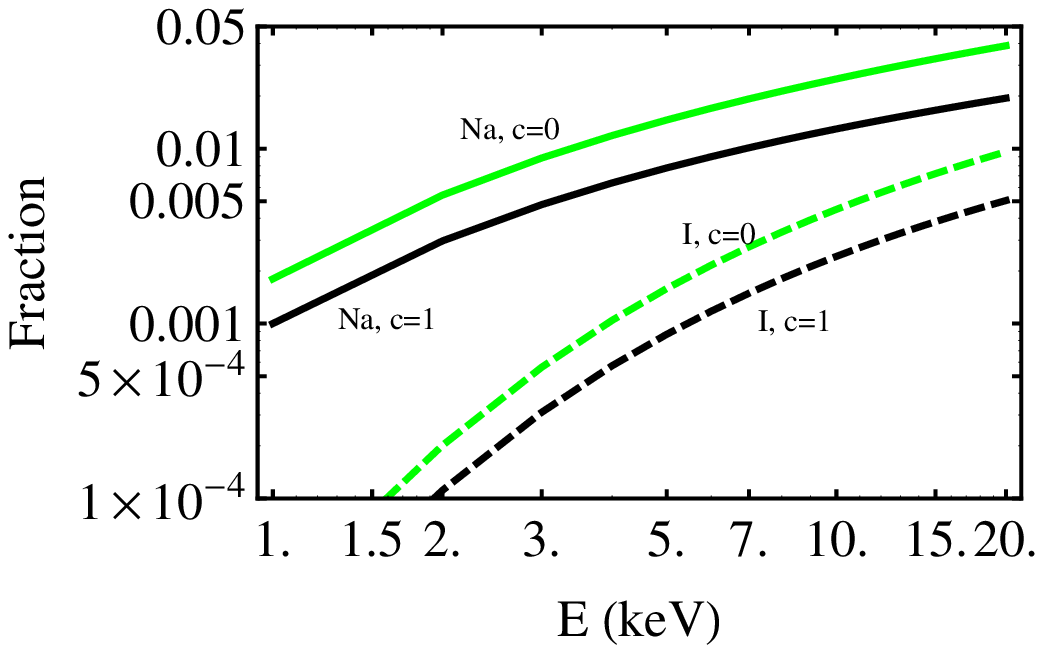}\\
  \vspace{-0.5cm}\caption{(Color online) Upper bounds to the (a) channeling fractions for single channels $\chi_i(E,\hat{\bf q})$ of Na recoils for axial (black lines) and planar (green/gray lines) channels with $c=1$, and (b) geometric channeling fraction $P_{\rm geometric}(E)$ of Na (solid lines) and I recoils (dashed lines) as a function of the recoil energy $E$ for $T=293$ K in with $c=0$ (green/gray) and $c=1$ (black), always without including dechanneling.}
  \label{Fraction-Na}
\end{center}
\end{figure}

Fig.~\ref{Fraction-Na}.a and  \ref{Fraction-Na}.b reproduced from  Ref.~\cite{BGGI}, show respectively upper bounds to some channeling fractions for single channels $\chi_i(E_R,\hat{\bf q})$ for Na recoils (with $c=1$) and geometric channeling  fraction of Na and I recoiling ions in a NaI crystal at room temperature  for  $1~{\rm keV}<E_R<20$ keV.  The parameter $c$ mentioned in the figures is a number that we expect to be between 1 and 2, which regulates the importance of temperature corrections  (for details see Ref.~\cite{BGGI}). The channeling fractions are typically smaller for larger values of $c$ thus setting $c=0$, which is an unrealistic value, we get  the largest upper bound to the channeling fractions that our calculations provide. In the figures we used $c=0$ and $c=1$. Notice also that the results in the figures do not take into account dechanneling effects which should also decrease the channeling fractions (we do not know how to properly take into account these effects with our analytic methods).

In this paper, we use the (upper bounds to the) channeling probability $\chi(E_R,\hat{\bf q})$  and the actual differential recoil spectrum  to compute the event rate, taking into account channeled and non-channeled recoils (see Section III, in particular Eqs.~\ref{prob} and \ref{def-Rate} and compare them with  Eq.~\ref{geometric-fraction}). We then use this rate to compute upper bounds to  the amplitude of the daily modulation due to channeling expected in NaI crystals. In Section IV, we examine the possibility that such a daily modulation  might be observable in the data accumulated by the DAMA collaboration.

\section{Angular distribution of recoil directions due to WIMPs}

Consider the WIMP-nucleus elastic collision for a WIMP of mass $m$ and a nucleus of mass $M$.
The 3-dimensional ``Radon transform" of the WIMP velocity distribution can be used to define the differential recoil spectrum as function of the recoil momentum $\vec{\bf q}$~\cite{Gondolo:2002}
\begin{equation}
\frac{dR}{ dE_R~d\Omega_q}= \frac{\rho \sigma_0 S(q)}{4\pi m \mu^2} \hat{f}_{\rm lab}\!\left( \frac{q}{2\mu}, \hat{\bf q} \right) ,
\label{eq: rate}
\end{equation}
where $E_R$ is the recoil energy,
$d\Omega_q=d\phi d\cos\theta$ denotes an infinitesimal solid angle around the recoil direction $\hat{\bf q}= \vec{\bf q}/q$, $q=|\vec{\bf q}|$ is the magnitude of the recoil momentum,  $\mu=mM/(m+M)$ is the reduced WIMP-nucleus mass, $q/2\mu= v_q$ is the minimum velocity a WIMP must have to impart a recoil momentum $q$ to the nucleus, or equivalently to deposit
a recoil energy $E_R = q^2 /2M$, $ \rho$ is the dark matter density in the solar neighborhood, $\sigma_0$ is the total scattering cross section of the WIMP  with a (fictitious) point-like nucleus, and $S(q)$ is the nuclear form factor normalized to 1.

We concentrate here on WIMPs with spin-independent interactions, for which $\sigma_0$ is usually written in terms of the
WIMP-proton cross section $\sigma_p$~\cite{Alenazi-Gondolo:2008}
\begin{equation}
\sigma_0=\frac{\mu^2}{\mu_p^2}A^2\sigma_p,
\label{sigma0}
\end{equation}
where $\mu_p=m m_p/(m+m_p)$ is the WIMP-proton reduced mass and $A$ is the atomic number of the nucleus.
We use the Helm form factor~\cite{Helm:1956}
\begin{equation}
S(q)=|F_{SI}(q)|^2=\left(\frac{3j_1(qR_1)}{qR_1}\right)^2 e^{-q^2 s^2},
\end{equation}
where
\begin{equation}
j_1(x)=\frac{\sin x}{x^2}-\frac{\cos x}{x}
\end{equation}
is the first kind spherical Bessel function, $R_1$ is an effective nuclear radius, and $s$ is the nuclear skin thickness. Following Duda, Kemper, and Gondolo ~\cite{Duda:2007} we set
\begin{equation}
R_1=\sqrt{c^2 + \frac{7}{3} \pi^2 a^2 - 5 s^2},
\end{equation}
and take $s\simeq0.9$ fm, $a\simeq0.52$ fm, and $c\simeq (1.23 A^{1/3} - 0.6)$ fm. These parameters have been chosen to match the numerical integration of  the Two-Parameter Fermi model of nuclear density~\cite{Duda:2007}.

The Maxwellian WIMP velocity distribution with respect to the Galaxy,  with dispersion $\sigma_v$ and truncated at the escape speed $v_{\rm esc}$  is given by~\cite{Gondolo:2002}
\begin{equation}
f_{\rm WIMP}({\bf v})=\frac{1}{N_{\rm esc} (2\pi \sigma_v^2)^{3/2}}\exp{\left[ -\frac{({\bf v}+{\bf V}_{\rm lab})^2}{2 \sigma_v^2} \right]},
\label{VelDist}
\end{equation}
for $|{\bf v}+{\bf V}_{\rm lab}|<v_{\rm esc}$, and zero otherwise, where
\begin{equation}
N_{esc}=\mathop{\rm erf}\left(\frac{v_{\rm esc}}{\sqrt{2}\sigma_v}\right)-\sqrt{\frac{2}{\pi}}\frac{v_{\rm esc}}{\sigma_v}\exp{\left[-\frac{v_{\rm esc}^2}{2\sigma_v^2} \right]}.
\label{Radon-transform}
\end{equation}
Here we are assuming the detector has a velocity $\textbf{V}_{\rm lab}$ with respect to the Galaxy (thus  $- \textbf{V}_{\rm lab}$ is the average velocity of the WIMPs with respect to the detector). ${\bf V}_{\rm lab}$ is defined in terms of the galactic rotation velocity ${\bf V}_{\rm {Gal Rot}}$ at the position of the Sun (or Local Standard of Rest (LSR) velocity), Sun's peculiar velocity ${\bf V}_{\rm {Solar}}$ in the LSR, Earth's translational velocity ${\bf V}_{\rm {Earth Rev}}$ with respect to the Sun, and the  velocity of Earth's rotation around itself ${\bf V}_{\rm {Earth Rot}}$ (see Appendix B),
\begin{equation}
{\bf V}_{\rm {lab}}={\bf V}_{\rm {Gal Rot}}+{\bf V}_{\rm {Solar}}+{\bf V}_{\rm {Earth Rev}}+{\bf V}_{\rm {Earth Rot}}.
\label{Vlab}
\end{equation}
In this paper we take $V_{\rm GalRot}$ either 220 km/s or 280 km/s, as reasonable low and high values (as done in Ref~\cite{Green-2010}),  which correspond to $V_{\rm lab}$ either 228.4 km/s or 288.3 km/s, respectively (see Appendix B for details). Ref.~\cite{Kuhlen} gives 100 km/s as the smallest estimate for the 1D velocity dispersion, which corresponds to a 3D dispersion  $\sqrt{3}$ times larger, i.e.  $\sigma_v=173$ km/s.  Thus here we take $\sigma_v$ either 173 km/s or 300 km/s~\cite{Gondolo:2002}.

In order to visualize the arrival directions of WIMPs, we will plot  $f_{\rm WIMP}(\hat{\bf v},v_q)$,  the number of WIMPs per solid angle  in the direction $\hat{\bf v}$ in several figures. If we limit ourselves to the WIMPs with speed higher than ${v_q}$, then
\begin{equation}
f_{\rm WIMP}(\hat{\bf v},v_q)=\int_{v_q}^{v_{\rm max}(\hat{\bf v})}{f_{\rm WIMP}({\bf v}) v^2 dv}.
\label{fWIMP}
\end{equation}
The upper limit of the integral in Eq.~\ref{fWIMP} is such that $|{\bf v}+{\bf V}_{\rm lab}|=v_{\rm esc}$ and depends on the direction  $\hat{\bf v}$, since $({\bf v}+{\bf V}_{\rm lab})^2=v^2+2v~ \hat{\bf v}.{\bf V}_{\rm lab} +V_{\rm lab}^2$,
\begin{equation}
v_{\rm max}(\hat{\bf v})=-\hat{\bf v}.{\bf V}_{\rm lab}+\sqrt{(\hat{\bf v}.{\bf V}_{\rm lab})^2-{\bf V}_{\rm lab}^2+v_{\rm esc}^2}~,
\end{equation}
and
\begin{equation}
f_{\rm WIMP}(\hat{\bf v},v_q)=\frac{\exp{\left(-\frac{V_{\rm lab}^2}{2 \sigma_v^2}\right)}}{N_{\rm esc} (2\pi \sigma_v^2)^{3/2}} \int_{v_q}^{v_{\rm max}(\hat{\bf v})}\exp{\left(\frac{-v^2}{2 \sigma_v^2}\right)} \exp{\left(\frac{-2v~\hat{\bf v}.{\bf V}_{\rm lab}}{2 \sigma_v^2}\right)} v^2 dv.
\end{equation}
This integral can be solved analytically and the result is in terms of error functions,
\begin{align}
f_{\rm WIMP}(\hat{\bf v},v_q)&=\frac{\exp{\left(-\frac{V_{\rm lab}^2}{2 \sigma_v^2}\right)}}{N_{\rm esc} (2\pi \sigma_v^2)^{3/2}}\left(\frac{\sigma_v}{2}\right)     \bigg\{\sqrt{2\pi}\left[(\hat{\bf v}.{\bf V}_{\rm lab})^2 + \sigma_v^2\right] \exp{\left(\frac{(\hat{\bf v}.{\bf V}_{\rm lab})^2}{2\sigma_v^2}\right)}\nonumber\\
&\left[\textrm{erf}\left(\frac{\hat{\bf v}.{\bf V}_{\rm lab}+v_{\rm max}(\hat{\bf v})}{\sqrt{2}\sigma_v}\right)-\textrm{erf}\left(\frac{\hat{\bf v}.{\bf V}_{\rm lab}+v_q}{\sqrt{2}\sigma_v}\right)\right]\nonumber\\
&+(2 \sigma_v)\bigg[\left(\hat{\bf v}.{\bf V}_{\rm lab}-v_{\rm max}(\hat{\bf v})\right) \exp{\left(-\frac{v_{\rm max}(\hat{\bf v})(2\hat{\bf v}.{\bf V}_{\rm lab} +v_{\rm max}(\hat{\bf v}))}{2\sigma_v^2}\right)}\nonumber\\
&+ \left(-\hat{\bf v}.{\bf V}_{\rm lab}+v_q\right) \exp{\left(-\frac{v_q(2\hat{\bf v}.{\bf V}_{\rm lab} +v_q)}{2\sigma_v^2}\right)} \bigg]\bigg\}.
\label{fTM}
\end{align}

\begin{figure}
\begin{center}
  \includegraphics[height=200pt]{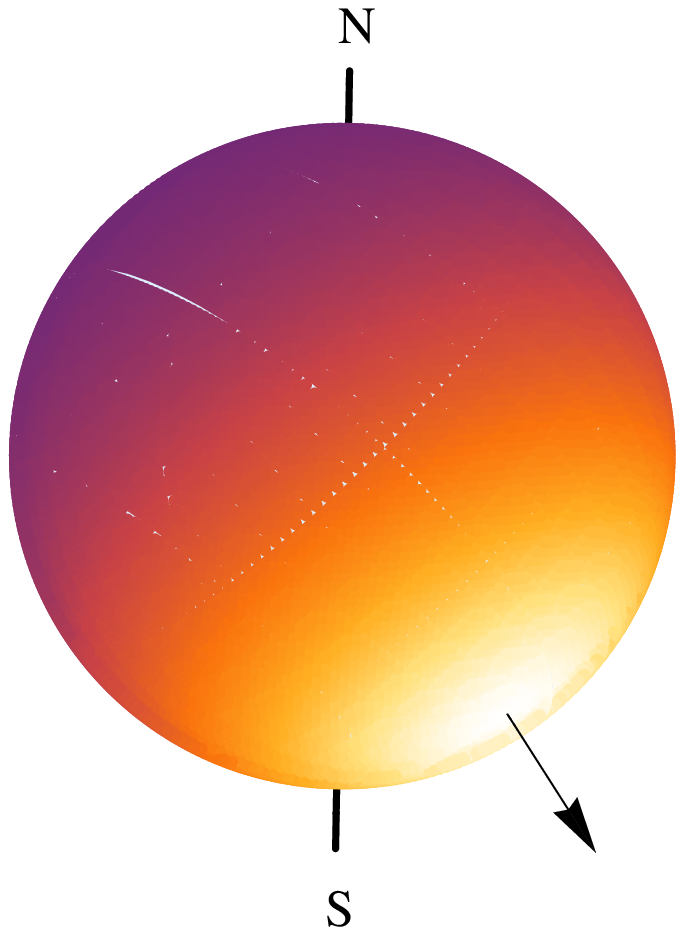}
  \includegraphics[height=200pt]{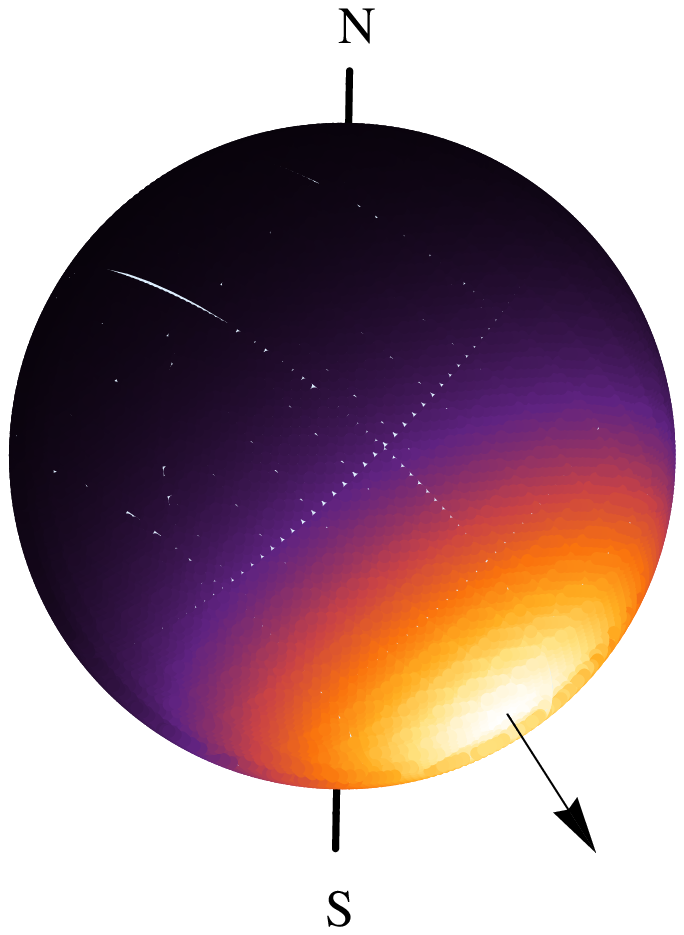}
  \includegraphics[height=30pt]{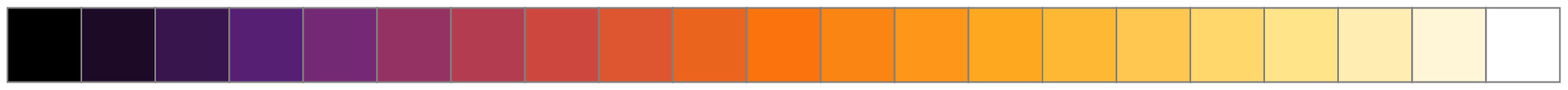}\\
  \vspace{-0.5cm}\caption{(Color online) WIMPs number density per solid angle $f_{\rm WIMP}(\hat{\bf v},v_q)$ (in Eq.~\ref{fTM})  for all WIMPs (namely  $v_q=0$) re-scaled to be a number between 0 (black) and 1 (white) plotted on the sphere of velocity directions $\hat{\bf v}$ using the HEALPix pixelization for   $V_{\rm {lab}}=288.3$ km/s and (a) $\sigma_v=300$ km/s and (b) $\sigma_v=173$ km/s. The arrow shows the direction of the average velocity of the WIMP wind,  $-{\bf V}_{\rm lab}$. The North and South celestial poles are also indicated. The color scale/grayscale shown in the horizontal bar between black and white corresponds to values between 0 and 1 in increments of 0.05.}%
  \label{Healpix-vq0}
\end{center}
\end{figure}

The maximum of $f_{\rm WIMP}(\hat{\bf v},v_q)$ happens when $\hat{\bf v}.{\bf V}_{\rm lab}=-V_{\rm lab}$, i.e. in the direction of the ``WIMP wind'' average velocity $-{\bf V}_{\rm lab}$.  Dividing $f_{\rm WIMP}(\hat{\bf v},v_q)$ by this maximum  we obtain a re-scaled distribution, a dimensionless number between 0 and 1, which we plot in  Fig.~\ref{Healpix-vq0} (see the color scale/grayscale in the figure) on the sphere  of  velocity directions $\hat{\bf v}$ using the HEALPix pixelization~\cite{HEALPix:2005} (see also Appendix B of Ref.~\cite{BGGI}) for all WIMPs, which amounts to taking $v_q=0$.  We took $V_{\rm lab}=288.3$ km/s, and $\sigma_v=300$ km/s or $\sigma_v=173$ km/s for Fig.~\ref{Healpix-vq0}.a or b respectively.

For a truncated Maxwellian WIMP velocity distribution with respect to the Galaxy, truncated at the escape speed $v_{\rm esc}$, the Radon-transform is~\cite{Gondolo:2002}
\begin{equation}
\hat{f}_{\rm lab}\!\left( \frac{q}{2\mu}, \hat{\bf q} \right)=\frac{1}{{N_{\rm esc}(2\pi \sigma_v^2)^{1/2}}}~{\left\{\exp{\left[-\frac{\left[ (q/2\mu) + \hat{\bf q} . {\bf V}_{\rm lab}\right]^2}{2\sigma_v^2}\right]}-\exp{\left[\frac{-v_{\rm esc}^2}{2\sigma_v^2}\right]}\right\}},
\label{fhatTM}
\end{equation}
 if $(q/2\mu) + \hat{\bf q} . {\bf V}_{\rm lab} < v_{\rm esc}$, and zero otherwise.

The presence of $\hat{\bf q} . {\bf V}_{\rm lab}$ means that in order to compute the differential rate we  need  to orient the nuclear recoil direction $\hat{\textbf{q}}$ with respect to ${\bf V}_{\rm lab}$.

The maximum of $\hat{ f}_{\rm lab} (\frac{q}{2 \mu},\hat{\textbf{q}})$ in Eq.~\ref{fhatTM} happens when $\hat{\bf q} . {\bf V}_{\rm lab}=-q/2\mu$, if $v_q=q/2 \mu < V_{\rm lab}$ (or in the direction of $-{\bf V}_{\rm lab}$ otherwise). Thus, we can re-scale $\hat{f}_{\rm lab}$ to obtain a dimensionless number between 0 and 1,
\begin{equation}
\hat{f}^{\rm re-scaled}_{\rm lab} ={\left\{\exp{\left[-\frac{\left[ (q/2\mu) + \hat{\bf q} . {\bf V}_{\rm lab}\right]^2}{2\sigma_v^2}\right]}-\exp{\left[\frac{-v_{\rm esc}^2}{2\sigma_v^2}\right]}\right\}}\bigg/\bigg(1-\exp{\left[\frac{-v_{\rm esc}^2}{2\sigma_v^2}\right]}\bigg).
\label{fhatTM-rescaled}
\end{equation}
 \begin{figure}
\begin{center}
  \includegraphics[height=200pt]{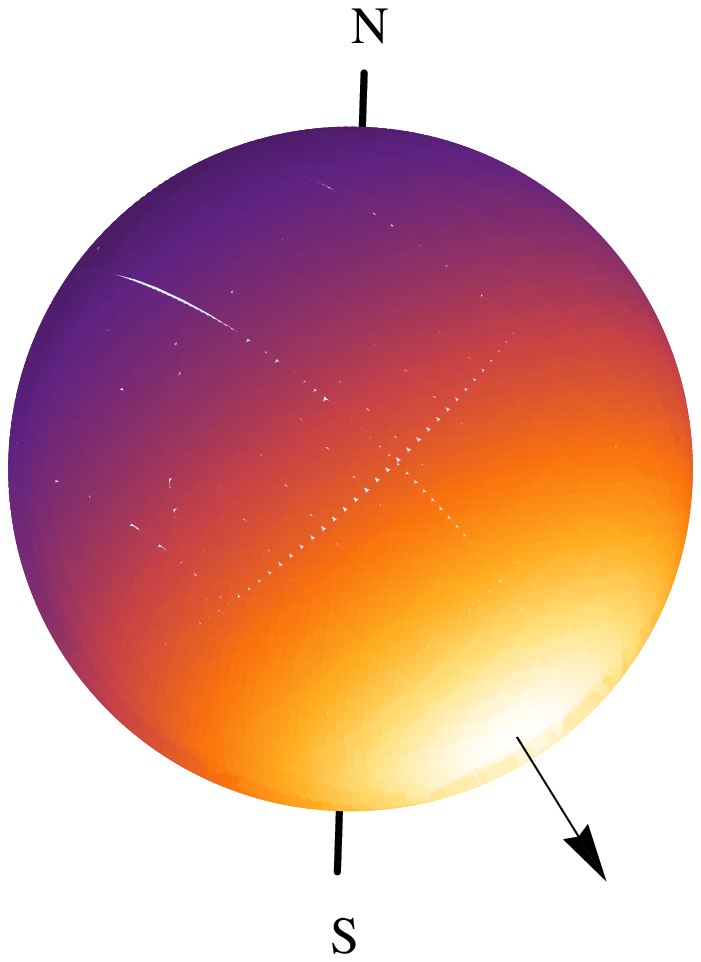}
  \includegraphics[height=200pt]{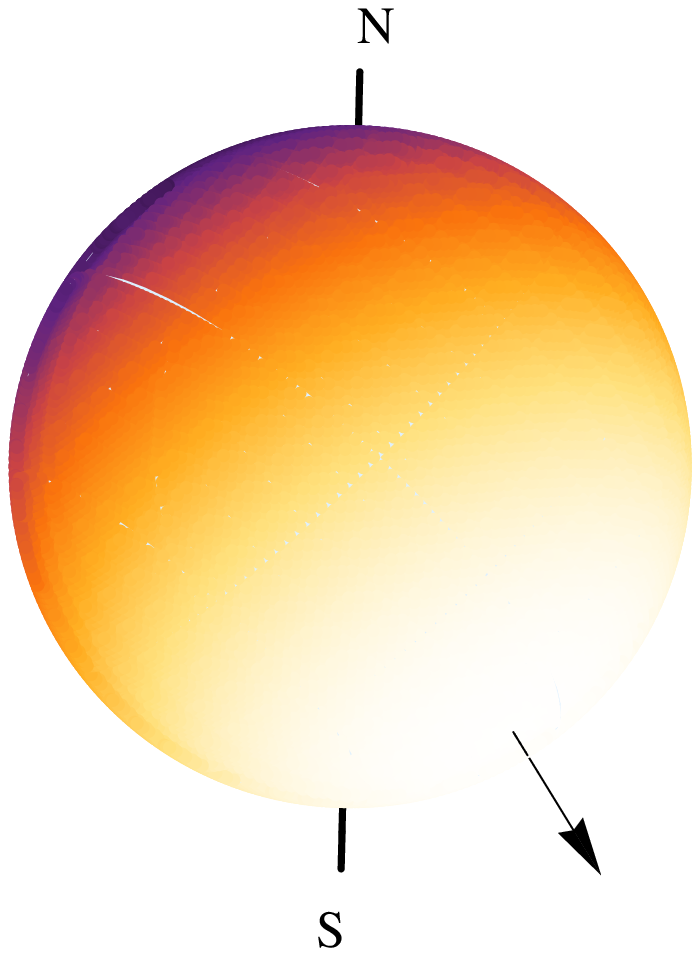}
  \includegraphics[height=30pt]{ColorBar.eps}\\
  \vspace{-0.5cm}\caption{(Color online) (a) $f_{\rm WIMP}(\hat{\bf v},v_q)$ (in Eq.~\ref{fTM}) re-scaled to be between 0 and 1 plotted on the sphere of velocity directions $\hat{\bf v}$ and (b) $\hat{f}_{\rm lab}$ (re-scaled as in Eq.~\ref{fhatTM-rescaled}) plotted on the sphere of recoil directions using the HEALPix pixelization for I recoils with  $E_R=10$ keV, $m=30$ GeV (thus $v_q=304.6$ km/s), $V_{\rm {lab}}=288.3$ km/s and $\sigma_v=300$ km/s. The arrow shows the direction of the average velocity of the WIMP wind,  $-{\bf V}_{\rm lab}$. The North and South celestial poles are also indicated. The color scale/grayscale shown in the horizontal bar corresponds to values between 0  (black) and  1 (white)  in intervals of 0.05. }
  \label{Healpix-1}
\end{center}
\end{figure}
  \begin{figure}
\begin{center}
  \includegraphics[height=200pt]{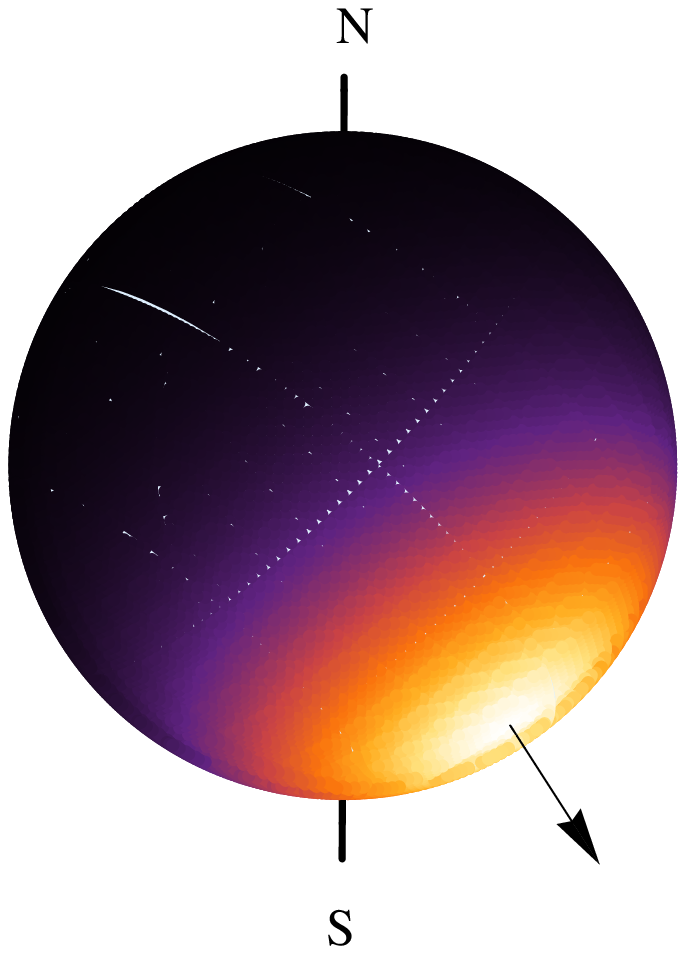}
  \includegraphics[height=200pt]{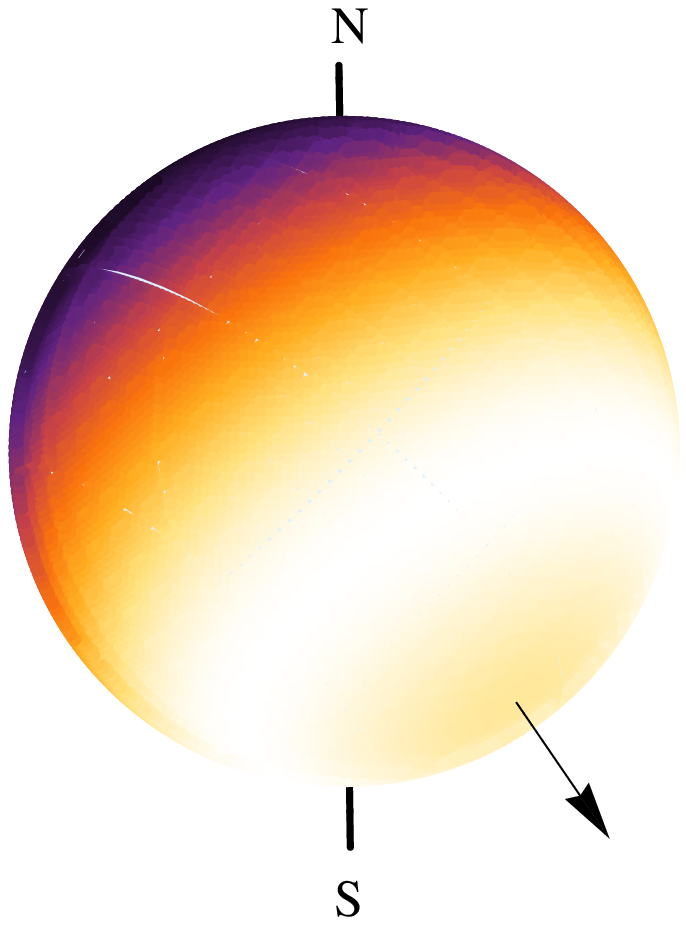}
  \includegraphics[height=30pt]{ColorBar.eps}\\
  \vspace{-0.5cm}\caption{(Color online) Same as Fig.~\ref{Healpix-1} but for Na recoils and assuming $m=60$ GeV (so $v_q=196.7$ km/s) and $\sigma_v=173$ km/s (and all other parameters the same).}
  \label{Healpix-2}
\end{center}
\end{figure}

In Figs.~\ref{Healpix-1} and~\ref{Healpix-2} we present side by side the WIMPs velocity distribution, for WIMPs which can generate a signal of a certain energy $E$, namely with speed above $v_q$ (left panels) and the Radon transform (right panels) of the recoils of energy $E$ that WIMP collisions produce.

 In Fig.~\ref{Healpix-1}.a  and b we respectively plot $f_{\rm WIMP}(\hat{\bf v},v_q)$ on the sphere  of WIMP velocity directions $\hat{\bf v}$  and $\hat{f}_{\rm lab}$ on the sphere  of recoil directions  (both re-scaled to be a number between 0 and 1)    using the HEALPix pixelization~\cite{HEALPix:2005} for I recoils assuming $V_{\rm lab}=288.3$ km/s, $E_R=10$ keV, $\sigma_v=300$ km/s and $m=30$ GeV. Fig.~\ref{Healpix-2}.a  and b show the same two distributions but for Na recoils and assuming $\sigma_v=173$ km/s and $m=60$ GeV (other parameters are the same).  The color scale/grayscale plotted on the spheres indicate different values  of the rescaled distributions: between 0  (black) and  1 (white)  in intervals of 0.05. In Fig.~\ref{Healpix-1} the minimum WIMP speed required is $v_q=304.6$ km/s (I recoils), and since $v_q >V_{\rm lab}$, the maximum value  of $\hat{f}^{\rm re-scaled}_{\rm lab}$, i.e. the maximum recoil rate, is in the direction of the ``WIMP wind''  average velocity, $-V_{\rm lab}$, which is shown with an arrow. In Fig.~\ref{Healpix-2} instead, $v_q=196.7$ km/s (Na recoils) and the maximum value  of $\hat{f}^{\rm re-scaled}_{\rm lab}$ occurs when $-\hat{\bf q} . {\bf V}_{\rm lab}=v_q$, i.e. when $\hat{\bf q}$ is at  an angle of $47^\circ$ of $-V_{\rm lab}$.

\section{Differential energy spectrum}

Let $p(E,E_R,\hat{\textbf{q}})dE$ be the probability that an energy $E$ is measured when a nucleus recoils in the direction $\hat{\textbf{q}}$ with initial energy $E_R$,  normalized so that
\begin{equation}
\int{p(E,E_R,\hat{\textbf{q}})dE}=1.
\end{equation}

With our analytic approach we cannot estimate the importance of dechanneling mechanisms, such as the presence of lattice imperfections, impurities or dopants. Thus we disregard dechanneling, and assume that a recoiling nucleus can only either be channeled, in which case the measured energy is the whole initial recoil energy $E=E_R$ (first term in the following equation) or not channeled, in which case the measured energy is $E= Q E_R$ (second term),
\begin{equation}
p(E,E_R,\hat{\textbf{q}})=\chi(E_R, \hat{\textbf{q}})\delta(E-E_R)+[1-\chi(E_R, \hat{\textbf{q}})]\delta(E-QE_R).
\label{prob}
\end{equation}
The first term accounts for the channeled (unquenched) events and the second term for the unchanneled (quenched) events, and $Q$ is the quenching factor.

Using Eq.~\ref{prob} the differential energy spectrum,
\begin{equation}
\frac{dR}{dE}=\int{\frac{dR}{dE_R d\Omega_q}p(E,E_R,\hat{\textbf{q}})d\Omega_q dE_R},
\label{def-Rate}
\end{equation}
can be written as
\begin{eqnarray}
\frac{dR}{dE}&=&\int{\left[\chi(E,\hat{\textbf{q}})\, \frac{dR}{dE_R d\Omega_q}\bigg|_{E_R=E} + [1-\chi(E/Q,\hat{\textbf{q}})]\,\frac{1}{Q} \, \frac{dR}{dE_R d\Omega_q}\bigg|_{E_R=E/Q} \right]d\Omega_q}\nonumber\\
&=&\frac{dR}{dE}\bigg|_{\rm U}
+\;\int{\left[\chi(E,\hat{\textbf{q}})\, \frac{dR}{dE_R d\Omega_q}\bigg|_{E_R=E} - \chi(E/Q,\hat{\textbf{q}})\frac{1}{Q} \, \frac{dR}{dE_R d\Omega_q}\bigg|_{E_R=E/Q} \right]d\Omega_q},
\end{eqnarray}
where the differential recoil spectrum with subindex ``U", which stands for ``Usual" (i.e. when channeling is not taken into account)  is
\begin{equation}
\frac{dR}{dE}\bigg|_{\rm U}=\int \frac{1}{Q}\frac{dR}{dE_Rd\Omega_q}\bigg|_{E_R=E/Q} d\Omega_q = \frac{1}{Q}\frac{dR}{dE_R}\bigg|_{E_R=E/Q}.
\end{equation}
Defining $\tilde{q} \equiv \sqrt{2EM}$ and using Eq.~\ref{eq: rate}, the measured differential rate becomes,
\begin{eqnarray}
\frac{dR}{dE}&=&\frac{dR}{dE}\bigg|_{\rm U}\;+\;\frac{\rho \sigma_0}{4\pi m \mu^2} \, \bigg[ S(\tilde{q}) \int{\chi(E,\hat{\textbf{q}}) \hat{f}_{\rm lab}\!\left( \frac{\tilde{q}}{2\mu}, \hat{\bf q} \right) d\Omega_q}\nonumber\\
&&- \frac{S(\tilde{q}/\sqrt{Q})}{Q} \int{\chi(E/Q,\hat{\textbf{q}}) \hat{f}_{\rm lab}\!\left( \frac{\tilde{q}}{2\mu \sqrt{Q}}, \hat{\bf q} \right) d\Omega_q } \bigg].
\end{eqnarray}
Inserting $\sigma_0$ from Eq.~\ref{sigma0} in the above equation with the usual value for the mean local halo density
$\rho=0.3~{\rm GeV/cm}^3$, we can write  the spin-independent detection rate of WIMPs in general  for a crystal that may contain more than one element
\begin{eqnarray}
\frac{dR}{dE}&=&\frac{dR}{dE}\bigg|_{\rm U}\;+\;1.306\times10^{-3}\frac{\text{events}}{\text{kg-day-keV}}\times \frac{\sigma_{44}}{4\pi m \mu_p^2} \sum_n C_n \, A_n^2 \, \bigg[ S(\tilde{q}) \int{\chi_n(E,\hat{\textbf{q}}) \hat{f}_{\rm lab}\!\left( \frac{\tilde{q}}{2\mu_n}, \hat{\bf q} \right) d\Omega_q}\nonumber\\
&&- \frac{S(\tilde{q}/\sqrt{Q_n})}{Q_n} \int{\chi_n(E/Q_n,\hat{\textbf{q}}) \hat{f}_{\rm lab}\!\left( \frac{\tilde{q}}{2\mu_n\sqrt{Q_n}}, \hat{\bf q} \right) d\Omega_q } \bigg],
\label{Rate}
\end{eqnarray}
where $\sigma_{44}$ is the WIMP-proton cross section in units of $10^{-44}\;{\text{cm}}^2$, $\mu_p$ and $m$ are in GeV and $\int{\hat{ f}_{\rm lab}d\Omega_q}$ is in ${(\text{km/s})}^{-1}$. The sum is over the nuclear species $n$  in a crystal, and $C_n$, $\chi_n$, $Q_n$ and $\mu_n$ are the  mass fraction, the channeling probability, the quenching factor and  the reduced WIMP-nucleus mass for  the element $n$, respectively.
For example, for NaI crystals,  as used in the DAMA experiment, we have
$C_{\rm Na}=M_{\rm Na}/(M_{\rm Na} + M_{\rm I})$  and
$C_{\rm I}={M_{\rm I}}/({M_{\rm Na} + M_{\rm I}})$, where $M_{\rm Na}$ and $M_{\rm I}$ are the atomic masses of Sodium and Iodine respectively.

 The integrals in Eq.~\ref{Rate} cannot be computed analytically. We integrate numerically by performing a Riemann sum once the sphere of directions has been divided using HEALPix~\cite{HEALPix:2005} (see also Appendix B of Ref.~\cite{BGGI}). HEALPix provides a convenient way of dividing the surface of a sphere into equal area sectors, and in our  papers~\cite{BGGI, BGG} we use it for the first time to compute integrals over directions.

  With the same notation, the usual rate is
\begin{equation}
\frac{dR}{dE}\bigg|_{\rm U} = 1.306\times10^{-3}\frac{\text{events}}{\text{kg-day-keV}}\times \frac{\sigma_{44}}{4\pi m \mu_p^2} \sum_n C_n \, A_n^2 \,  \, \bigg[\frac{S(\tilde{q}/\sqrt{Q_n})}{Q_n} \int{\hat{f}_{\rm lab}\!\left( \frac{\tilde{q}}{2\mu_n\sqrt{Q_n}}, \hat{\bf q} \right) d\Omega_q } \bigg].
\end{equation}

\section{Daily Modulation in NaI Crystals}

We present here the daily modulation amplitude due to channeling expected in NaI crystals for several WIMP masses and Na or I recoil energies. We assume that WIMPs have a truncated Maxwellian velocity distribution as in Eq.~\ref{VelDist} with $v_{\rm esc}=650$ km/s. We use the upper bounds to  channeling fractions for single channels $\chi_i(E_R,\hat{\bf q})$ given in  Ref.~\cite{BGGI}. We take $T=293$ K, the temperature of the DAMA experiment.

The spin-independent detection rate of WIMPs given in Eq.~\ref{Rate} has a time dependence through the Radon transform $\hat{ f}_{\rm lab}$. Notice that $\hat{ f}_{\rm lab}$ (see Eq.~\ref{fhatTM}) changes during a day through the $(\hat{\bf q} . {\bf V}_{\rm lab})$ factor appearing in the exponent  and the dependence of ${\bf V}_{\rm lab}$ on ${\bf V}_{\rm {EarthRot}}$ (see Eq.~\ref{Vlab}). The expression showing the time dependence of $\hat{\bf q} . {\bf V}_{\rm lab}$ is given in Eq.~\ref{qdotVlab} (in Appendix  B).  During a day, ${\bf V}_{\rm {EarthRev}}$ which is responsible for the annual modulation changes too. Thus the rate does not return to exactly the same value after one day. For the cases we present in this paper, this difference is less than 10\% of the total modulation amplitude in a day, and we did not correct for this effect.

\subsection{Relative Modulation Amplitudes}

Here we show  the  signal rate as function of time during a particular arbitrary Solar day (September 25, 2010). We define  the relative  signal modulation amplitude $A_s$ (taking into account the signal only) in terms of the maximum and minimum daily signal rate $R_s$ as
 \begin{equation}
 A_s= \frac{R_{s {\rm -max}}-R_{s{\rm -min}}}{R_{s{\rm-max}}+R_{s{\rm-min}}}.
 \end{equation}
The total relative modulation amplitude $A_T$ is defined  in terms of the maximum $R_{T{\rm-max}}$ and minimum $R_{T{\rm -min}}$ total daily rates  as
 \begin{equation}
 A_T= \frac{R_{T {\rm -max}}-R_{T{\rm -min}}}{R_{T{\rm-max}}+R_{T{\rm-min}}}.
 \end{equation}
The total rate consists of signal plus background,  $R_T=R_s+R_b$. Assuming that there is no daily modulation in the background,  $R_{T{\rm -max}}-R_{T{\rm -min}}=$ $R_{s{\rm-max}}-R_{s{\rm -min}}$, and $A_T$ is related to $A_s$ as
\begin{equation}
A_T=A_s(R_s/R_T),
\label{AT}
\end{equation}
where the average total rate due to signal and background is $R_T= (R_{T{\rm-max}}+R_{T{\rm-min}})/2$ and the average rate due to the signal alone is $R_s= (R_{s{\rm-max}}+R_{s{\rm-min}})/2$.

\begin{figure}
\begin{center}
  \includegraphics[height=190pt]{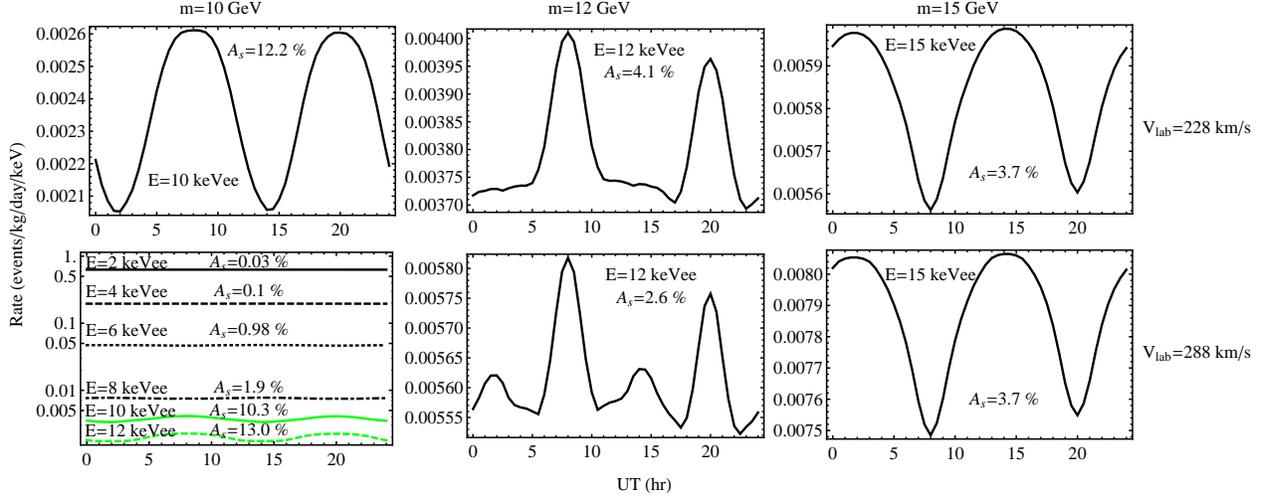}
  \vspace{-0.5cm}\caption{Signal rate (in events/kg-day-keVee) as function of the Universal Time (UT) during 24 hours for $m=10$ GeV, 12 GeV and 15 GeV for different energies. The parameters used are $\sigma_v=300$ km/s, $Q_{\rm Na}=0.2$, $Q_{\rm I}=0.09$, $\sigma_p=2 \times 10^{-40} \textrm{cm}^2$, $c=1$ for temperature effects, a crystal temperature of $T=293$ K and   $V_{\rm {lab}}=228.4$ km/s (top row) or 288.3 km/s (bottom row).}
	\label{FigureTable}
\end{center}
\end{figure}

Exploring the parameter space of WIMP mass and WIMP-proton cross section for different recoil energies we find that the relative modulation amplitudes $A_s$ can be large,  even more than 10\%  for some combination of parameters. We explored the range of WIMP masses  from a few GeV to hundreds of GeV for recoil energies between 2 keV and a few MeV. We show some examples in  Fig.~\ref{FigureTable}, where we plot the  signal rate (in events/kg/day/keVee)   as function of  the Universal Time (UT) during 24 hours.  We find that the largest $A_s$ happen when the signal is only due to channeling. This happens when there are no WIMPs in the galactic halo with large enough kinetic energy to provide the observed energy if the recoil is not channeled. The observed energies for which the rate is only due to channeling depend on the quenching factors $Q$, which are not well known. The  smaller values of $Q$ make channeling more important so we take  $Q_{\rm Na}=0.2$~\cite{Hooper-Collar}  for Na and  the usual $Q_{\rm I}=0.09$ for I.

\subsection{Statistical Significance}

The detectability  of a particular amplitude of daily modulation depends on the exposure and background of a particular experiment. The former DAMA/NaI and the DAMA/LIBRA experiments (which we refer collectively as  the DAMA experiment) have a very large cumulative exposure, 1.17 ton $\times$ year. However even with this large exposure, we find that the daily modulations we predict are not observable.  To observe the daily modulation,  the  total number of events  $N_T$  ($N_s$ signal plus $N_b$ background events) over the duration of the experiment  should be divided into two bins,  the ``high-rate''  bin with $N_{T {\rm -max}}$ events and the ``low-rate'' bin with $N_{T {\rm -min}}$ events,  so that  $N_T=N_{T {\rm -max}}+N_{T {\rm -min}}$.  For the daily modulation to be observable at, say, the 3$\sigma$ level one should have
\begin{equation}
N_{T {\rm -max}}-N_{T {\rm -min}}= A_T N_T > 3\sigma \simeq 3 \sqrt{N_T/2},
\label{RateCond}
\end{equation}
where  $\sigma^2 \simeq N_T/2$ because, with a small modulation, on average $N_{T {\rm -max}} \simeq N_{T {\rm -min}} \simeq N_T/2$.  In principle there are other errors associated with identifying the  ``high-rate'' and ``low-rate''  bins which we do not include here. Thus we are underestimating the errors.

 If the detector exposure is $M T$ in kg-day and we take bins of width $\Delta E$  in keVee, then  $N_{T {\rm -max}} = R_{T {\rm -max}} M T \Delta E /2$,  $N_{T {\rm -min}} = R_{T {\rm -min}} M T \Delta E /2$, $N_T =R_T M T~ \Delta E$ and $N_s =R_s M T \Delta E$,  where the rates are in events/kg-day-keVee. Thus  $(N_s/N_T)=(R_s/R_T)$ and using Eq.~\ref{AT}, $A_T=A_s(N_s/N_T)$.  Thus the condition in Eq.~\ref{RateCond} becomes $A_s N_s > 3 \sqrt{N_T/2}$ which implies
\begin{equation}
N_s^2/N_T > 9 / (2 A_s^2),
\end{equation}
or
\begin{equation}
R_s^2/R_T > 9 /(2 A_s^2 M T~ \Delta E).
\label{RateCond2}
\end{equation}

 The total rate of the DAMA experiment at low energies $4~{\rm keVee}<E<10$ keVee is $R_T \simeq 1 ~{\rm events/kg/day/keVee}$~\cite{DAMA-bckg}. This rate is much larger than the signal rates we predict  and is, therefore, dominated by background. With this value of $R_T$, Eq.~\ref{RateCond2} becomes
\begin{equation}
R_s^2 A_s^2 > {\frac{9}{2  M T ~\Delta E~{\rm kg~day~keVee}}}.
\label{RateCond3}
\end{equation}
We choose here a bin $\Delta E \simeq 1$ keVee, narrow enough to assume the signal rate to be constant in it and compatible with the energy resolution of DAMA. The energy resolution of DAMA  is  $\sigma_E(E)=(0.448~{\rm keVee})\sqrt{E/{\rm keVee}}+(0.0091) E \simeq 1$ keVee at low energies~\cite{Bernabei:2008}. We consider the significance of the highest signal-to-noise energy bin  that we found through inspection. With the cumulative exposure of DAMA, the condition in Eq.~\ref{RateCond3} for relative daily modulation amplitudes $A_s$ observable at 3$\sigma$ is
\begin{equation}
R_s~A_s>3.2 \times 10^{-3}~{\rm events/kg/day/keVee},
\end{equation}
or
\begin{equation}
R_{s {\rm -max}}-R_{s {\rm -min}} > 6.4 \times 10^{-3}~{\rm events/kg/day/keVee}.
\label{RateCond4}
\end{equation}
For observability at the $n \sigma$ level we should multiply the right-hand side of Eq.~\ref{RateCond4} by $(n/3)$. Even the largest relative daily modulations we find, shown in Fig.~\ref{FigureTable}, are not observable in the DAMA data according to Eq.~\ref{RateCond4}.

The examples which we show here are for small WIMP masses and recoil energies. For large masses the value of $\sigma_p$ must be chosen in the region of the cross section and mass plane where XENON10/100 and CDMS impose $\sigma_p$ to be smaller by four orders of magnitude than for light WIMPs. This amounts to corresponding smaller signal rates and ($R_{s {\rm -max}}-R_{s {\rm -min}}$) differences. For small WIMP masses and large energies, $v_q$ is large and there are no WIMPs with the speed required for Na or I recoils. Thus, only small WIMP masses and recoil energies result in high modulation amplitudes.

Fig.~\ref{FigureTable} shows the  signal rate  during 24 hours for three different WIMP masses $m=10$ GeV, 12 GeV and 15 GeV and different  energies $E$ between 2 and 15 keVee. The other relevant parameters are $\sigma_v=300$ km/s,  $\sigma_p=2 \times 10^{-40} \textrm{cm}^2$ (close to the DAMA and CoGeNT regions~\cite{Hooper-Collar, Savage:2010, CDMS:2011}), $c=1$, $T=293$ K  and two values of $V_{\rm lab}$, 228.4 km/s (top row) and 288.3 km/s (bottom row). Recent bounds, e.g. those from XENON100~\cite{Xenon100:2011}, impose smaller values of $\sigma_p$. In any event,  changes in $\sigma_p$ are easy to take into account because $A_s$ is independent of $\sigma_p$ and  the rate is just proportional  to it, $R_s \sim \sigma_p$.

  We found the relative amplitude $A_s$ to be as large as 12\% in the examples shown in Fig.~\ref{FigureTable}, but even those large values are not observable according to Eq.~\ref{RateCond4} (even at the 1$\sigma$ level). With the choice of $V_{\rm lab}=228.4$ km/s (top row of Fig.~\ref{FigureTable}) we get a signal rate difference $R_{s {\rm -max}}-R_{s {\rm -min}}$ of $0.56 \times 10^{-3}$ events/kg/day/keVee for $m=10$ GeV and $E=10$ keVee (in this case $v_q=454.8$ km/s  and 790.5 km/s  for channeled Na and I recoils, respectively), $3.17 \times 10^{-4}$ events/kg/day/keVee for $m=12$ GeV and $E=12$ keVee (which corresponds to $v_q=441.6$ km/s and 732.9 km/s for Na and I channeled recoils, respectively), and $4.25 \times 10^{-4}$ events/kg/day/keVee for $m=15$ GeV and $E=15$ keVee (for which $v_q=430.6$ km/s and 670.6 km/s for Na and I channeled recoils, respectively). With the choice of $V_{\rm lab}=288.3$ km/s (bottom row of Fig.~\ref{FigureTable}),  $R_{s {\rm -max}}-R_{s {\rm -min}}$ is $0.77 \times 10^{-3}$ events/kg/day/keVee for $m=10$ GeV and $E=10$ keVee (one of the energies shown), $2.95 \times 10^{-4}$ events/kg/day/keVee for $m=12$ GeV and $E=12$ keVee, and $0.58 \times 10^{-5}$ events/kg/day/keVee for $m=15$ GeV and $E=15$ keVee. Because the minimum WIMP speeds  $v_q$  are large in these examples, a smaller velocity dispersion  of the WIMP distribution leads to smaller rates (since a smaller amount of WIMPs  have velocities larger than $v_q$). So the signal rate difference $R_{s {\rm -max}}-R_{s {\rm -min}}$ is even smaller for smaller values of  $\sigma_v$.

 The left-bottom panel of Fig.~\ref{FigureTable} shows the signal rate as function of UT for $m=10$ GeV and  $V_{\rm lab}=288.3$ km/s for several energies between 2 keVee and 12 keVee. The rate decreases but $A_s$ increases with increasing energy and the best conditions for observability happen at some energy where neither the rate nor $A_s$ are very small. The rates for low energies between 2 keVee and 6 keVee are dominated by the usual  (i.e. non-channeled) rate and the daily modulation is due purely to the change in WIMP kinetic energy in the lab frame as the Earth rotates around itself. The rates for energies above 8 keVee (green/gray lines) are purely due to channeling, i.e. the usual rate is zero. For intermediate energies, 6 keVee to 8 keVee, the usual and channeled rates both contribute and thus the daily modulation is due to both the channeling effect and the daily change in the usual rate. For $E=2$, 4, 6, 8, 10 and 12 keVee, the values of $R_{s {\rm -max}}-R_{s {\rm -min}}$ given in  events/kg/day/keVee are  respectively  $4.3 \times 10^{-4}$, $0.5 \times 10^{-3}$, $0.92 \times 10^{-3}$, $2.8 \times 10^{-4}$, $0.77 \times 10^{-3}$  and $0.52 \times 10^{-3}$. Notice that for all the energies shown the difference in rate is similar, but the largest $A_s$ values happen at energies above 8 keVee, for which the rate is only due to channeling. The channeling daily modulation amplitude increases as  the ratio of the velocity dispersion to the average speed of the WIMPs that contribute to the signal (i.e. with $v>v_q$) decreases. This ratio is small and thus $A_s$ large for large values of $v_q$. Notice that the phase of the modulation due to channeling depends on the orientation of the crystal with respect to the Galaxy and the phase of the modulation in the usual rate does not, which would allow to distinguish both effects, if they were observable. The case of $m=10$ GeV and $E=6$ keVee has the largest rate difference,  but is not observable at 3$\sigma$ according to Eq.~\ref{RateCond4} (not even at the 1$\sigma$ level). Choosing $\sigma_p=4 \times 10^{-40} \textrm{cm}^2$ (still within the DAMA allowed region but not compatible with the recent XENON100 result) results in a rate difference of $1.84 \times 10^{-3}$ events/kg/day/keVee for this case which would not be observable even at the 1$\sigma$ level.

\begin{figure}
\begin{center}
  \includegraphics[height=144pt]{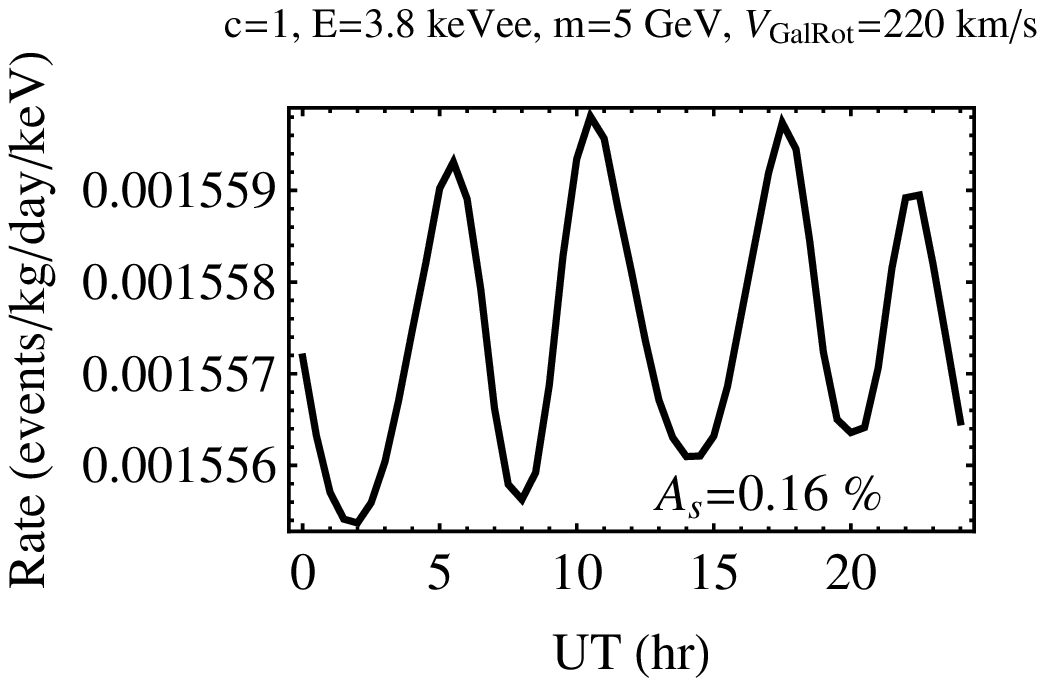}
  \includegraphics[height=144pt]{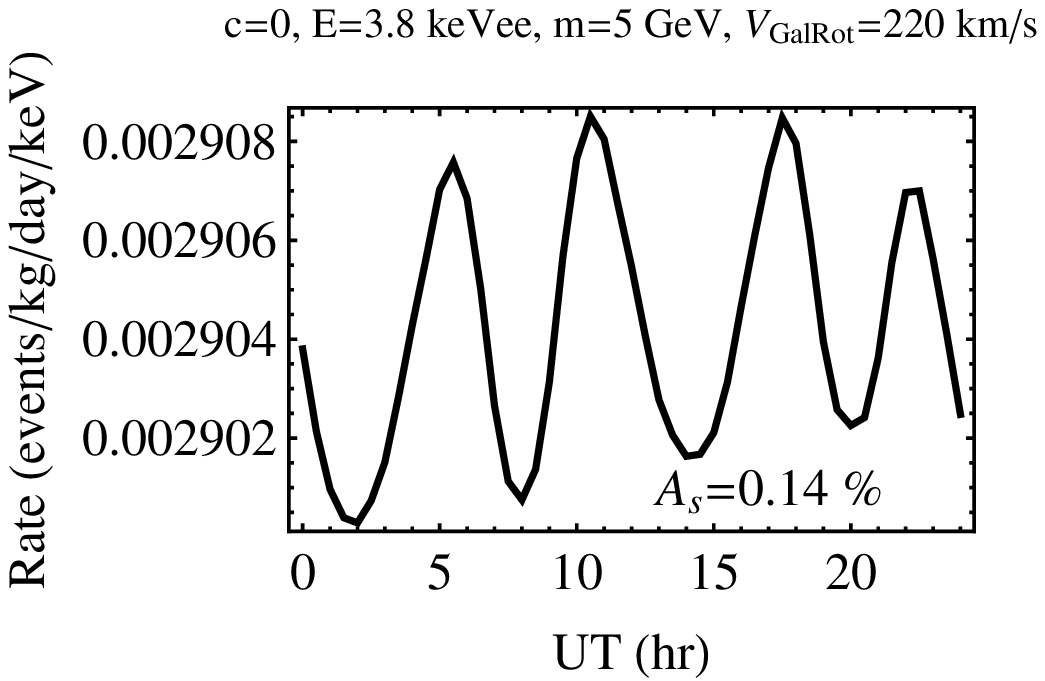}\\
  \vspace{-0.5cm}\caption{Signal rate as function of UT during 24 hours for $E=3.8$ keVee and $m=5$ GeV, with $V_{\rm {lab}}=228.4$ km/s, $\sigma_v=300$ km/s, $\sigma_p=2 \times 10^{-40} \textrm{cm}^2$, and $Q_{\rm Na}=0.2$, $Q_{\rm I}=0.09$ for (a) $c=1$ and (b) $c=0$. The daily modulation is not observable in both cases.}
	\label{T293-e3m5}
\end{center}
\end{figure}

 Finally, we would like to compare our results with those obtained in Ref.~\cite{Avignone:2008cw} by Creswick {\it et al}. They  found  a relative daily modulation amplitude $A_s=$0.85\% (their definition of amplitude differs by a factor of 2 from ours, so they quote 1.7\%) for  5 GeV WIMP mass  and  3.8 keVee measured energy (in which case $v_q=471.2$ km/s and 936.6 km/s for  channeled Na and I recoils, respectively. There are no WIMPs with the speed required for I recoils, thus only Na recoils are possible). In order to compare our calculation with theirs, we compute the signal event rate as function of time  for $c=1$, $T=293$ K (temperature corrections are not included in the calculation of Creswick {\it et al.}) and choosing all the other parameters very close to those used in Ref.~\cite{Avignone:2008cw}, i.e.  $V_{\rm {lab}}=228.4$ km/s and $\sigma_v=300$ km/s. A WIMP mass of 5 GeV is outside the region of parameter space compatible with the annual modulation reported by DAMA~\cite{Savage:2010}. Since $A_s$ does not depend on $\sigma_p$, we choose an arbitrary value of $\sigma_p=2 \times 10^{-40} \textrm{cm}^2$ to plot the signal rate as a function of UT (the upper bound given by  TEXONO and CoGeNT~\cite{TEXONO:2009} is five times larger, $\sigma_p < 1 \times 10^{-39} \textrm{cm}^2$). Our result  is shown in  Fig.~\ref{T293-e3m5}.a. We find $A_s=$0.16\% ($R_{s {\rm -max}}-R_{s {\rm -min}}=4.4 \times 10^{-6} ~{\rm events/kg/day/keVee}$). Even when we consider the extreme choice of $c=0$ to compute temperature effects (an unrealistic value for which the channeling fractions are larger) with the same parameters, we get $A_s=0.14\%$. This case is shown in Fig.~\ref{T293-e3m5}.b.

\subsection{Future Prospects for DAMA and other Experiments}

The daily modulation might be detectable in other experiments with smaller background or WIMP halo components with a smaller dispersion  such as streams or  a thick disk. The amplitude of the daily modulation increases as the WIMP velocity distribution is narrower i.e. for larger values of the average velocity and smaller values of the velocity dispersion of the detectable WIMPs (which is not $\sigma_v$), i.e. those with velocity larger than $v_q$.  This is easy to understand since as the dispersion increases more channels are available for channeling of the recoiling ions. In the limit in which  the velocity distribution would be isotropic with respect to the detector, the daily rotation would not introduce any difference in the rate due to channeling.  Having a large relative signal modulation amplitude $A_s$  is not sufficient for observability. In Eq.~\ref{RateCond4} what is important is $(A_s~ R_s)=(R_{s {\rm -max}}-R_{s {\rm -min}})/2$. However, the condition in Eq.~\ref{RateCond4}  was derived considering the total rate in the DAMA experiment, which is dominated by background. For an experiment where the background is negligible, i.e. $R_T=R_s+R_b \simeq R_s$, we can derive a different observability condition (at the 3$\sigma$ level) from Eq.~\ref{RateCond2},
\begin{equation}
R_s~A_s^2= A_s~(R_{s {\rm -max}}-R_{s {\rm -min}})/2> 9 /(2 M T~ \Delta E).
\label{RateCond-NoBckg}
\end{equation}
This condition might be  easier to satisfy in future experiments.

 One could ask which is the maximum level of total rate with the current DAMA exposure that would be needed to make the signal daily modulation observable. Inserting the current exposure of DAMA (1.17 ton  year) in Eq.~\ref{RateCond2}, we have
\begin{equation}
{\left(A_s~R_s\right)} ^2/R_T > 1.05 \times 10^{-5}~{\rm events/kg/day/keVee},
\end{equation}
which using  $A_s R_s=(R_{s {\rm -max}}-R_{s {\rm -min}})/2$, becomes
\begin{equation}
R_T < \frac{\left(R_{s {\rm -max}}-R_{s {\rm -min}}\right)^2}{4.2 \times 10^{-5} ~{\rm events/kg/day/keVee}}.
\end{equation}
Even in the case with the highest rate difference we found, i.e. $R_{s {\rm -max}}-R_{s {\rm -min}}=0.98 \times 10^{-3}$ events/kg/day/keVee (the $m=10$ GeV, $E=6$ keVee, $V_{\rm lab}=288.3$ km/s  example shown in the bottom-left panel of Fig.~\ref{FigureTable}) observability would require
\begin{equation}
R_T < 0.023  ~{\rm events/kg/day/keVee},
\label{maxR_T}
\end{equation}
roughly $1/40$ of what is now.

We could ask instead what exposure would be needed with the current total rate in the DAMA experiment  to make the daily modulation observable. Setting $R_T \simeq 1$ events/kg/day/keVee in Eq.~\ref{RateCond2}, we obtain
\begin{equation}
\frac{M T \Delta E}{{\rm (events/kg/day/keVee)}} > \frac{9}{2 \left(A_s~R_s \right)^2}=  \frac{18}{\left(R_{s {\rm -max}}-R_{s {\rm -min}}\right)^2}  .
\end{equation}
Again, for the case with the highest rate difference we found  ($m=10$ GeV, $E=6$ keVee and $V_{\rm lab}=288.3$ km/s) and with $\Delta E \simeq$ 1 keVee we would require an exposure 40 times larger,
\begin{equation}
M T  > 51.3 ~{\rm ton ~year}.
\end{equation}

\begin{figure}
\begin{center}
  \includegraphics[height=144pt]{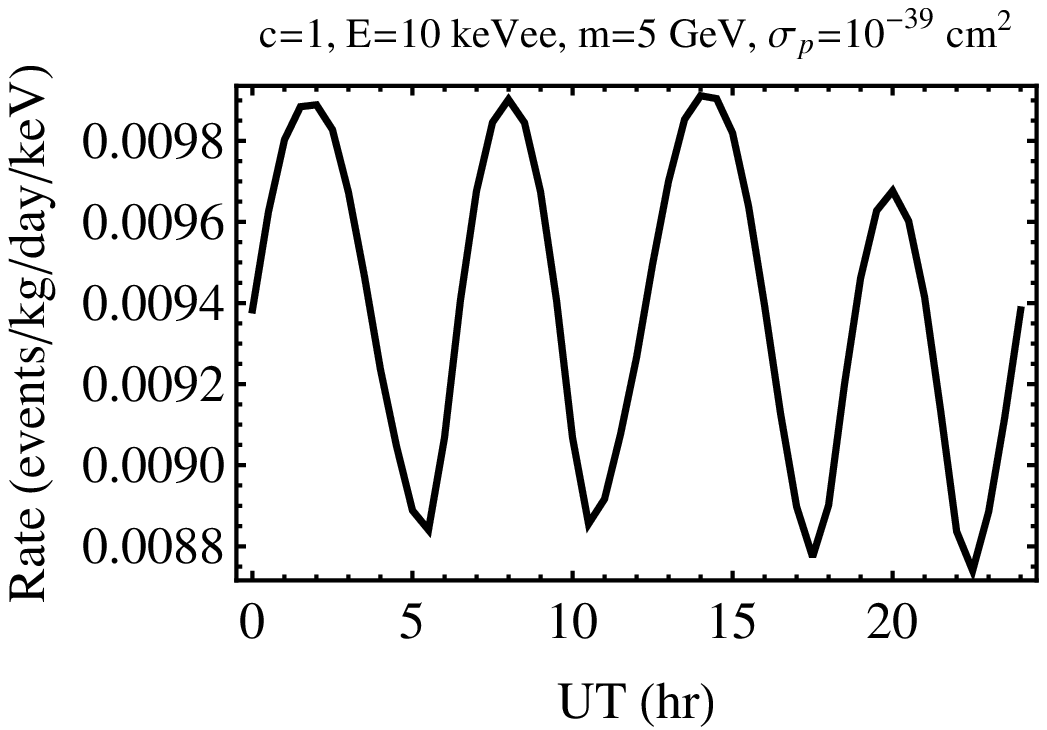}
  \includegraphics[height=144pt]{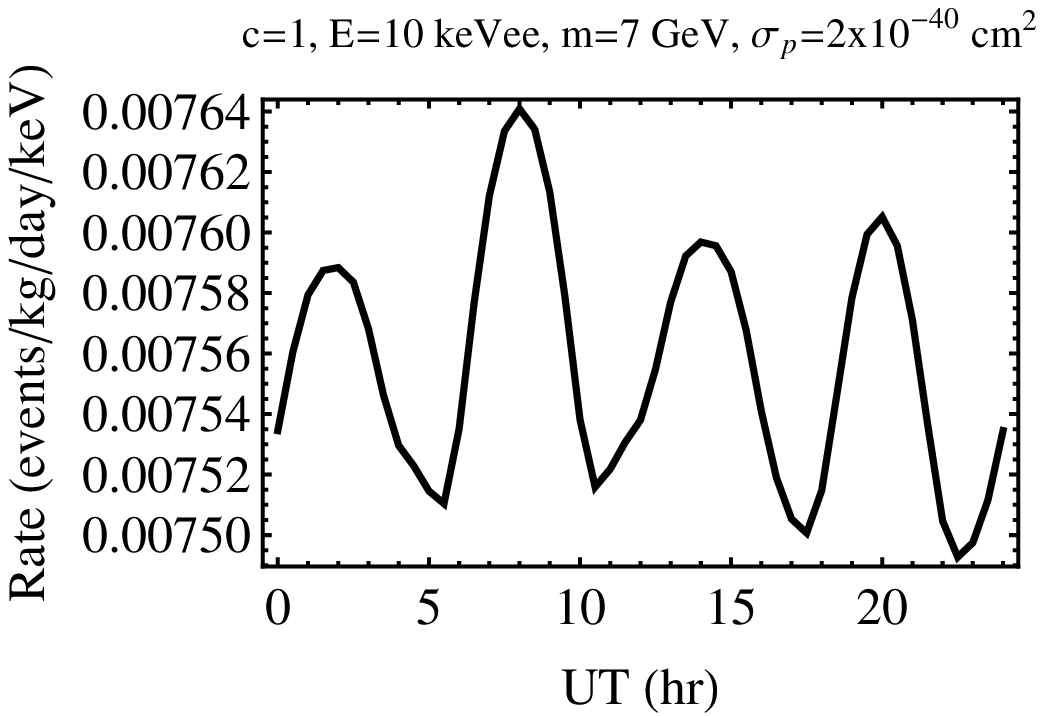}\\
  \vspace{-0.5cm}\caption{Signal rate as function of UT during 24 hours for a solid Ne detector operating at $T=23$ K at Gran Sasso for $E=10$ keVee, $Q=0.25$, $c=1$, $\sigma_v=300$ km/s, $V_{\rm {lab}}=228.4$ km/s and for (a) $m=5$ GeV and $\sigma_p=10^{-39} \textrm{cm}^2$, and (b) $m=7$ GeV and $\sigma_p=2 \times 10^{-40} \textrm{cm}^2$.}
	\label{NeRate}
\end{center}
\end{figure}

 We have computed the daily modulation due to channeling in other material such as Ge, solid Xe and solid Ne, and we find that  it will be very difficult to observe.  For light WIMPs the cross section can be larger than for heavier ones without violating experimental bounds,  $\sigma_p=10^{-39} \textrm{cm}^2$~\cite{TEXONO:2009} and this favors the detection of the daily modulation.  We find that  for a WIMP mass $m=5$ GeV the daily modulation due to channeling may be observable in solid Ne if the signal would be above threshold and  assuming no background. The geometric  channeling fraction reaches a maximum at around 10 keV for solid Ne~\cite{BGG}, thus the largest modulation amplitude happens at that energy. For example for a solid Ne detector operating at 23 K at Gran Sasso, for $E=10$ keV, assuming $Q_{\rm Ne} =0.25$~\cite{Tretyak}, $c=1$ and with velocity distribution parameters $\sigma_v=300$ km/s and $V_{\rm {lab}}=228.4$ km/s we find $R_s A_s^2=3.68 \times 10^{-5}$ events/kg/day/keVee. Using Eq.~\ref{RateCond-NoBckg} we find that the exposure needed to observe this modulation at 3$\sigma$ is $MT=0.33$ ton year. For the same parameters but for $m=7$ GeV and $\sigma_p=2 \times 10^{-40} \textrm{cm}^2$ (parameters compatible with the possible dark matter signal found by  CoGeNT and with DAMA according to Ref.~\cite{Hooper}), we find $R_s A_s^2=7.2 \times 10^{-7}$ events/kg/day/keVee, and the exposure needed is $MT=17.1$ ton year. The usual rate is zero in both cases, and the modulation is just due to channeling.  The signal rate during 24 hours and the required exposures for the two cases are shown in Fig.~\ref{NeRate} and Table~\ref{table:Ne}, respectively.
\begin{table}[ht]
\caption{Observability in solid Ne detector} 
\centering
\begin{tabular}{c c c} 
\hline\hline 
Case & $\sigma_p$ ($\textrm{cm}^2$) & $MT$ (ton year) \\ [0.5ex] 
\hline 
$m=5$ GeV & ~~$10^{-39}$~~ & 0.33 \\ 
$m=7$ GeV & ~~$2 \times 10^{-40}$~~ & 17.1 \\ [1ex] 
\hline 
\end{tabular}
\label{table:Ne}
\end{table}

We intend to further explore the  observability of a daily modulation in future experiments for different halo models in future work.

\begin{acknowledgments}
G.G. and N.B.  were supported in part by the US Department of Energy Grant
DE-FG03-91ER40662, Task C.  P.G. was  supported  in part by  the NFS
grant PHY-0756962 at the University of Utah.
\end{acknowledgments}

\appendix
\addappheadtotoc

\section{Crystal Orientation}

 We need to orient the crystal with respect to the laboratory. We define a reference frame fixed with the laboratory and orient its axes so that the $xy$ plane is horizontal, the $x$-axis points North, the $y$-axis points West, and the $z$-axis points to the zenith. We denote its unit coordinate vectors as $\hat{\cal N}$, $\hat{\cal W}$ and $\hat{\cal Z}$, respectively. We also define the crystal frame with $X,Y,Z$ cartesian axes fixed with the crystal. The unit coordinate vectors of the crystal frame are $\hat{\mathbf{X}}$, $\hat{\mathbf{Y}}$ and $\hat{\mathbf{Z}}$.

We now want to connect the laboratory frame to the crystal frame. Let the standard orientation correspond to the configuration in which $\hat{\mathbf{X}}=\hat{\cal N}$, $\hat{\mathbf{Y}}=\hat{\cal W}$, and $\hat{\mathbf{Z}}=\hat{\cal Z}$. We start with the crystal in the standard orientation, and we turn it into any other orientation $\hat{\mathbf{X}}$, $\hat{\mathbf{Y}}$, $\hat{\mathbf{Z}}$. In this new orientation, each of the unit coordinate vectors of the crystal frame can be written in terms of unit coordinate vectors of the lab frame,
\begin{eqnarray}
\hat{\mathbf{X}}&=&\alpha_X~\hat{\mathbf{\mathcal{N}}}+\beta_X~\hat{\mathbf{\mathcal{W}}}+\gamma_X~\hat{\mathbf{\mathcal{Z}}},\nonumber\\
\hat{\mathbf{Y}}&=&\alpha_Y~\hat{\mathbf{\mathcal{N}}}+\beta_Y~\hat{\mathbf{\mathcal{W}}}+\gamma_Y~\hat{\mathbf{\mathcal{Z}}},\nonumber\\
\hat{\mathbf{Z}}&=&\alpha_Z~\hat{\mathbf{\mathcal{N}}}+\beta_Z~\hat{\mathbf{\mathcal{W}}}+\gamma_Z~\hat{\mathbf{\mathcal{Z}}},
\label{Crystal-Lab}
\end{eqnarray}
where $\alpha_i$, $\beta_i$ and $\gamma_i$ are the ``direction cosines'' between the two sets of cartesian coordinates of the lab and crystal frames, for $i=X,Y,Z$. For example, the coordinate vector $\hat{\mathbf{X}}$ of the crystal has a particular angle with each of the lab frame coordinate vectors $\hat{\cal N}$, $\hat{\cal W}$, $\hat{\cal Z}$. Let $a_X$ be the angle between $\hat{\mathbf{X}}$ and $\hat{\cal N}$, $b_X$ the angle between $\hat{\mathbf{X}}$ and $\hat{\cal W}$, and $c_X$ the angle between $\hat{\mathbf{X}}$ and $\hat{\cal Z}$. The direction cosines of the unit vector $\hat{\mathbf{X}}$ are given by,
\begin{eqnarray}
\alpha_X &\equiv& \cos a_X =\hat{\mathbf{X}}\cdot\hat{\cal N},\nonumber\\
\beta_X &\equiv& \cos b_X =\hat{\mathbf{X}}\cdot\hat{\cal W},\nonumber\\
\gamma_X &\equiv& \cos c_X =\hat{\mathbf{X}}\cdot\hat{\cal Z}.
\end{eqnarray}
We can find the direction cosines for $\hat{\mathbf{Y}}$ and $\hat{\mathbf{Z}}$ unit vectors in a similar way. From these definitions it follows that $\alpha_i ~\alpha_j+\beta_i~\beta_j+\gamma_i~\gamma_j=\delta_{ij}$ where $i,j=X,Y,Z$. We prefer using direction cosines over Euler angles because the direction cosines can easily be measured for any known orientation of a crystal in a laboratory, whereas it may be difficult to specify the Euler angles.

Eq.~\ref{Crystal-Lab} gives the transformation from the lab frame to the crystal frame. We can also find the lab coordinate vectors in terms of the crystal coordinate vectors,
\begin{eqnarray}
\hat{{\bf {\cal N}}}&=&\alpha_X~\hat{{\bf X}}+\alpha_Y~\hat{{\bf Y}}+\alpha_Z~\hat{{\bf Z}},\nonumber\\
\hat{{\bf {\cal W}}}&=&\beta_X~\hat{{\bf X}}+\beta_Y~\hat{{\bf Y}}+\beta_Z~\hat{{\bf Z}},\nonumber\\
\hat{{\bf {\cal Z}}}&=&\gamma_X~\hat{{\bf X}}+\gamma_Y~\hat{{\bf Y}}+\gamma_Z~\hat{{\bf Z}}.
\label{Lab-Crystal}
\end{eqnarray}
In the results we show in this paper, we took $\alpha_X=\beta_Y=\gamma_Z=1$ and all the other $\alpha_i$, $\beta_i$ and $\gamma_i$ equal to zero. Choosing a different orientation for the crystal does not change the average rate, but $A_s$ may change by a factor of 2 for NaI depending on the orientation of the crystal. The observability condition is still not satisfied.

\subsection{Lab to equatorial transformation}

 To connect the laboratory frame to the equatorial coordinate frame, we recall the definition of the geocentric equatorial inertial (GEI) frame: its origin is at the center of the Earth, its $x_e$-axis points in the direction of the vernal equinox, its $y_e$-axis points to the point on the celestial equator with right ascension 90$^\circ$ (so that the cartesian frame is right-handed), and its $z_e$-axis points to the north celestial pole. We denote its unit coordinate vectors as $\hat{\bf x}_e$,  $\hat{\bf y}_e$, and  $\hat{\bf z}_e$. We want to find the transformation formulas from the laboratory frame to the GEI frame.

This transformation can be achieved by two successive rotations. The first rotation is by an angle of $(90^\circ-\lambda_{\rm lab})$ counterclockwise about the laboratory $y$-axis to align the new $x' y'$ plane with the plane of the celestial equator. Here $\lambda_{\rm lab}$ is the latitude of the laboratory in degrees, with northern latitudes taken as positive and southern latitudes taken as negative. With this rotation, the new $z'$-axis points to the north celestial pole. The second rotation is by an angle $(15t_{\rm lab}+180)$ degrees clockwise about the new $z'$-axis to bring the $x'$-axis in the direction of the vernal equinox. Here $t_{\rm lab}$ is the laboratory Local  Apparent Sidereal Time (LAST) in hours (the LAST is the hour angle of the vernal equinox at the location of the laboratory). One has
 \begin{equation}
 t_{\rm lab} = t_{\rm GAST} + l_{\rm lab}/15,
 \end{equation}
 where $t_{\rm GAST}$ is the Greenwich Apparent Sidereal Time (GAST) in hours and $l_{\rm lab}$ is the longitude in degrees measured positive in the eastward direction (e.g.\ $l_{\rm lab}=+110^\circ$ for 110$^\circ$ E and $l_{\rm lab}=-110^\circ$ for 110$^\circ$ W).

The current local apparent sidereal time for any specified longitude $l_{\rm lab}$ can be computed online, for example on the website of the US Naval Observatory at http://tycho.usno.navy.mil/ sidereal.html (accessed Sept 19, 2010).   As an alternative, one can use the following formula \cite{Hapgood,USNO89} for the Greenwich mean sidereal time (which differs from the Greenwich apparent sidereal time by less than 1.2 seconds, completely negligible for our purposes),
\begin{align}
t_{\rm GAST} = (101.0308 + 36000.770 \, T_0 + 15.04107 \, {\rm UT})/15,
\label{tGAST}
\end{align}
where
\begin{align}
T_0 = \frac{ \lfloor {\rm MJD} \rfloor - 55197.5}{36525.0}.
\end{align}
Here ${\rm UT}$ is the Universal Time in hours, $\lfloor {\rm MJD} \rfloor $ is the integer part of the modified Julian date (MJD), which is the time measured in days from 00:00 UT on 17 November 1858 (Julian date 2400000.5). Note that $T_0$ is the time in Julian centuries (36525 days) from 12:00 UT on 1 January 2010 to the previous midnight. At 12:00 UT on 1 January 2010, the Julian date is 2455198, and the MJD is 55197.5. Also the the $15.04107/15$ in Eq.~\ref{tGAST} corrects from solar time (UT) to sidereal time. Sidereal day is shorter than Solar day by 3.9 minutes.  In this paper, all our results are computed for the particular arbitrary day of 25 September 2010, for which $T_0=0.00729637$.

Note also that UT is different from coordinated Universal Time (UTC) which is the time scale usually used for data recording. UTC is atomic time adjusted by an integral number of seconds to keep it within 0.6 s of UT. For our purposes the difference between UT and UTC is negligible.

Taking into account the two rotations explained above, one can find the transformation equations of the unit vectors,
\begin{align}
\hat{\bf x}_e & = -\cos(t^\circ_{\rm lab}) \left[ \sin(\lambda_{\rm lab}) \hat{\cal N} -\cos(\lambda_{\rm lab}) \hat{\cal Z} \right] + \sin(t^\circ_{\rm lab}) \hat{\cal W}, \nonumber\\
\hat{\bf y}_e & = -\sin(t^\circ_{\rm lab}) \left[ \sin(\lambda_{\rm lab}) \hat{\cal N} -\cos(\lambda_{\rm lab}) \hat{\cal Z}\right] - \cos(t^\circ_{\rm lab}) \hat{\cal W}, \nonumber\\
\hat{\bf z}_e & = \cos(\lambda_{\rm lab}) \hat{\cal N} + \sin(\lambda_{\rm lab}) \hat{\cal Z},
\label{eq:labeq3}
\end{align}
where $t^\circ_{\rm lab}=15 t_{\rm lab}$ is the laboratory LAST converted to degrees.

As a check, for a laboratory on the equator at local sidereal time 0, i.e.\ $\lambda_{\rm lab}=0^\circ$ and $t^\circ_{\rm lab}=0^\circ$, one has $\hat{\bf x}_e = \hat{\cal Z}$, $\hat{\bf y}_e = -\hat{\cal W}$, and $\hat{\bf z}_e = \hat{\cal N}$; six sidereal hours later at the same laboratory, i.e.\ $\lambda_{\rm lab}=0^\circ$ and $t^\circ_{\rm lab}=90^\circ$, one has $\hat{\bf x}_e = \hat{\cal W}$, $\hat{\bf y}_e = \hat{\cal Z}$, and $\hat{\bf z}_e = \hat{\cal N}$; for a laboratory at the South Pole ($\lambda_{\rm lab}=-90^\circ$), using the direction of the Greenwich meridian in place of the "North" axis $\hat{\cal N}$ so that the local sidereal time at the South Pole by convention coincides with the Greenwich sidereal time, one has $\hat{\bf x}_e = \hat{\cal N}$, $\hat{\bf y}_e = -\hat{\cal W}$, and $\hat{\bf z}_e = -\hat{\cal Z}$ at $t^\circ_{\rm lab}=0^\circ$ and $\hat{\bf x}_e = \hat{\cal W}$, $\hat{\bf y}_e = \hat{\cal N}$, and $\hat{\bf z}_e = -\hat{\cal Z}$ at $t^\circ_{\rm lab}=90^\circ$ . All of these are correctly given by Eq.~\ref{eq:labeq3}.

The formulas in Eq.~\ref{eq:labeq3} can be inverted, and the transformation from the equatorial frame to the lab frame is achieved:
\begin{align}
\hat{\cal N}&=-\sin(\lambda_{\rm lab})\left[\cos (t^\circ_{\rm lab}) \hat{\bf x}_e +\sin (t^\circ_{\rm lab})\hat{\bf y}_e\right]+\cos(\lambda_{\rm lab})\hat{\bf z}_e, \nonumber\\
\hat{\cal W}&=\sin (t^\circ_{\rm lab})\hat{\bf x}_e-\cos (t^\circ_{\rm lab})~\hat{\bf y}_e, \nonumber\\
\hat{\cal Z}&=\cos(\lambda_{\rm lab}) \left[\cos (t^\circ_{\rm lab}) \hat{\bf x}_e +\sin (t^\circ_{\rm lab}) \hat{\bf y}_e\right]+\sin(\lambda_{\rm lab})\hat{\bf z}_e.
\label{Equit-Lab}
\end{align}
 The latitude and longitude of Gran Sasso are $\lambda_{\rm lab}=42.45^\circ$ and $l_{\rm lab}=13.7^\circ$, respectively.

\begin{figure}
\begin{center}
  \includegraphics[height=247pt]{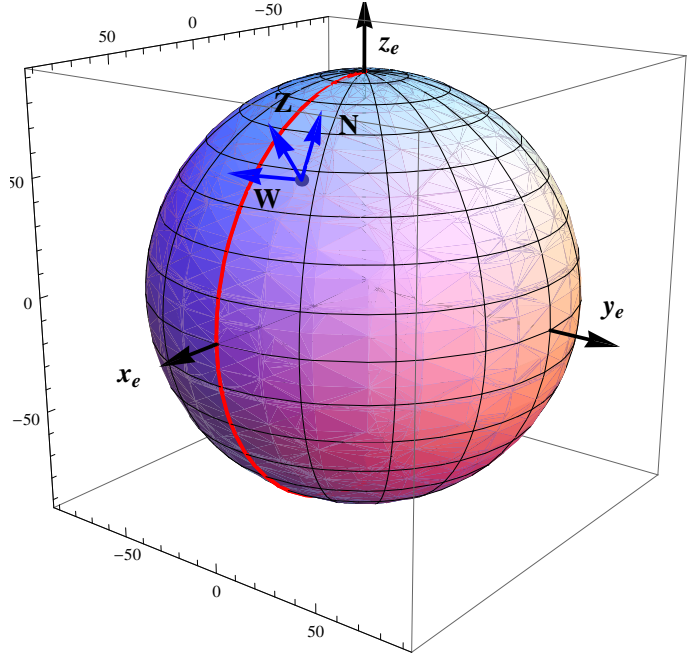}
 \caption{(Color online) Earth's sphere in the equatorial frame ($\hat{\bf x}_e$,$\hat{\bf y}_e$,$\hat{\bf z}_e$) specified with black arrows. The laboratory frame (N,W,Z) specified with blue/dark gray arrows is also shown.}
	\label{Earth}
\end{center}
\end{figure}

Fig.~\ref{Earth} shows the laboratory frame ($\hat{\cal N}$, $\hat{\cal W}$, $\hat{\cal Z}$) and the equatorial coordinate frame ($\hat{\bf x}_e$,$\hat{\bf y}_e$,$\hat{\bf z}_e$) plotted on the Earth's sphere at $UT=0$ using Eq.~\ref{Equit-Lab}.

\subsection{Equatorial to galactic transformation}

To connect the equatorial frame to the galactic coordinate frame, we recall the definition of the galactic coordinate system: its origin is at the position of the Sun, its $x_g$-axis points towards the galactic center, its $y_g$-axis points in the direction of the galactic rotation, and its $z_g$-axis points to the north galactic pole.

For the epoch of January 1950.0 the transformation from the equatorial frame ($\hat{\mathbf{x}}_e, \hat{\mathbf{y}}_e, \hat{\mathbf{z}}_e$) to the galactic frame ($\hat{\mathbf{x}}_g, \hat{\mathbf{y}}_g, \hat{\mathbf{z}}_g$) is given by~\cite{gal-equat}:
\begin{align}
\hat{{\bf x}}_g&=\hat{{\bf x}}_e~(-0.06699)+\hat{{\bf y}}_e~(-0.8728)+\hat{{\bf z}}_e~(-0.4835),\nonumber\\
\hat{{\bf y}}_g&=\hat{{\bf x}}_e~(0.4927)+\hat{{\bf y}}_e~(-0.4503)+\hat{{\bf z}}_e~(0.7446),\nonumber\\
\hat{{\bf z}}_g&=\hat{{\bf x}}_e~(-0.8676)+\hat{{\bf y}}_e~(-0.1883)+\hat{{\bf z}}_e~(0.4602).
\label{Equit-Gal}
\end{align}
The transformation from the galactic frame to the equatorial frame is given by
\begin{eqnarray}
\hat{\mathbf{x}}_e&=&\hat{\mathbf{x}}_g~(-0.06699)+\hat{\mathbf{y}}_g~(0.4927)+\hat{\mathbf{z}}_g~(-0.8676),\nonumber\\
\hat{\mathbf{y}}_e&=&\hat{\mathbf{x}}_g~(-0.8728)+\hat{\mathbf{y}}_g~(-0.4503)+\hat{\mathbf{z}}_g~(-0.1884),\nonumber\\
\hat{\mathbf{z}}_e&=&\hat{\mathbf{x}}_g~(-0.4835)+\hat{\mathbf{y}}_g~(0.7446)+\hat{\mathbf{z}}_g~(0.4602).
\label{Gal-Equit}
\end{eqnarray}
The change of Eqs.~\ref{Equit-Gal} and \ref{Gal-Equit} from the epoch of January 1950.0 to 25 September 2010 is small and would not affect the final results in this paper.

\section{Laboratory motion}

The velocity of the lab  with respect to the center of the Galaxy can be divided into four components (as in Eq.~\ref{Vlab}): ${\bf V}_{\rm {Gal Rot}}$, ${\bf V}_{\rm {Solar}}$, ${\bf V}_{\rm {Earth Rev}}$ and ${\bf V}_{\rm {Earth Rot}}$.

We take $V_{\rm {Gal Rot}}=220$ km/s or 280 km/s~\cite{Green-2010}, $V_{\rm Solar}=18$ km/s~\cite{Schoenrich-2010}, $V_{\rm {Earth Rev}}=29.8$ km/s and $V_{\rm {Earth Rot}}=(0.465102 ~{\rm km/s}) \cos \lambda_{\rm lab}$, where $\lambda_{\rm lab}$ is the latitude of the lab. Values of $V_{\rm {Gal Rot}}=220$ km/s or 280 km/s results in $V_{\rm lab}=228.4$ km/s or 288.3 km/s, respectively (see Appendix B.5 for the equation of ${\bf V}_{\rm {lab}}$). Thus, ${\bf V}_{\rm {lab}}$ is dominated by the galactic rotation velocity.

We need to compute $\hat{\mathbf{q}} \cdot{\bf V}_{{\rm lab}}$, where $\hat{\mathbf{q}}$ is given in the crystal reference frame ($\hat{\mathbf{q}}=q_X ~\hat{{\bf X}} +q_Y ~\hat{{\bf Y}} +q_Z ~\hat{{\bf Z}}$). Therefore, we need to also write ${\bf V}_{\rm {lab}}$ in the crystal frame. We have,
\begin{equation}
\hat{\bf q} \cdot{\bf V}_{\rm {lab}}=\hat{\bf q} \cdot{\bf V}_{\rm {Gal Rot}}+\hat{\bf q} \cdot{\bf V}_{\rm {Solar}}+\hat{\bf q} \cdot{\bf V}_{\rm {Earth Rev}}+\hat{\bf q} \cdot{\bf V}_{\rm {Earth Rot}}.
\label{qdotV}
\end{equation}
We will compute each term on the right-hand side of Eq.~\ref{qdotV} individually.

\subsection{Galactic rotation}

The velocity of the galactic rotation ${\bf V}_{\rm {Gal Rot}}$ is defined in the galactic reference frame,
\begin{equation}
{\bf V}_{\rm {Gal Rot}}=V_{\rm {Gal Rot}} \hat{{\bf y}}_g,
\label{GalacticRot}
\end{equation}
where $V_{\rm {Gal Rot}}$ is the galactic rotation speed (i.e. the local circular speed), and $\hat{{\bf y}}_g$ is in the direction of the galactic rotation. Following Ref.~\cite{Green-2010} , we take $V_{\rm {Gal Rot}}=220$ km/s or 280 km/s. Using the conversions in Eq.~\ref{Equit-Gal}, we can write $\hat{{\bf y}}_g$ in the equatorial reference frame in terms of ($\hat{{\bf x}}_e$,$\hat{{\bf y}}_e$,$\hat{{\bf z}}_e$). Then, we use Eq.~\ref{eq:labeq3} to transform from the equatorial frame to the lab frame ($\hat{\cal N},\hat{\cal W},\hat{\cal Z}$), and finally we use Eq.~\ref{Lab-Crystal} to transform from the lab frame to the crystal frame ($\hat{{\bf X}}, \hat{{\bf Y}},\hat{{\bf Z}}$).

Thus, we can use Eq.~\ref{Lab-Crystal} to write ${\bf V}_{\rm {Gal Rot}}$ in terms of the crystal frame coordinates, and compute $\hat{\mathbf{q}} \cdot{\bf V}_{\rm {Gal Rot}}$,
\begin{align}
\hat{\mathbf{q}} \cdot{\bf V}_{\rm {Gal Rot}}&=q_X V_{\rm {Gal Rot},X}+q_Y V_{\rm {Gal Rot},Y}+q_Z V_{\rm {Gal Rot},Z}.
\end{align}
We have
\begin{align}
\hat{\mathbf{q}} \cdot{\bf V}_{\rm {Gal Rot}}&=V_{\rm {Gal Rot}}\bigg\{\bigg(\left[-0.4927\cos(t^\circ_{\rm lab}) +0.4503\sin(t^\circ_{\rm lab}) \right]\sin(\lambda_{\rm lab})+0.7446\cos(\lambda_{\rm lab})\bigg)\nonumber\\
&\left(\alpha_X q_X+\alpha_Y q_Y+\alpha_Z q_Z\right)+\bigg(0.4927\sin(t^\circ_{\rm lab})+0.4503\cos(t^\circ_{\rm lab}) \bigg)\left(\beta_X q_X+\beta_Y q_Y+\beta_Z q_Z\right)\nonumber\\
&+\bigg(\left[0.4927\cos(t^\circ_{\rm lab}) -0.4503\sin(t^\circ_{\rm lab}) \right]\cos(\lambda_{\rm lab})+0.7446\sin(\lambda_{\rm lab})\bigg)\nonumber\\
&\left(\gamma_X q_X+\gamma_Y q_Y+\gamma_Z q_Z \right)\bigg\}.
\label{qdotGalacticRot}
\end{align}
Eq.~\ref{qdotGalacticRot} has a time dependence through $t^\circ_{\rm lab}$ and would be responsible for any daily modulation in the rate.

\subsection{Solar motion}

The velocity of the Sun's motion in the galactic rest frame is,
\begin{equation}
{\bf V}_{\rm {Solar}}=U \hat{{\bf x}}_g + V \hat{{\bf y}}_g +W \hat{{\bf z}}_g,
\label{Solar}
\end{equation}
where $(U,V,W)_\odot=(11.1, 12.2, 7.3)$ km/s~\cite{Schoenrich-2010}. Using Eq.~\ref{Equit-Gal}, we can transform from the galactic frame to the equatorial frame, and using Eq.~\ref{eq:labeq3} we can transform from the equatorial frame to the lab frame. Then we can use Eq.~\ref{Lab-Crystal} to write ${\bf V}_{\rm {Solar}}$ in terms of the crystal frame coordinates.

Thus, we can compute $\hat{\mathbf{q}} \cdot{\bf V}_{\rm {Solar}}$ as
\begin{align}
\hat{\bf q} \cdot{\bf V}_{\rm {Solar}}&=\bigg(\big[(1.066~\textrm{km/s})\cos(t^\circ_{\rm lab})+(16.56~\textrm{km/s})\sin(t^\circ_{\rm lab})\big]\sin(\lambda_{\rm lab})+(7.077 ~\textrm{km/s})\cos(\lambda_{\rm lab})\bigg)\nonumber\\
&(\alpha_X q_X+\alpha_Y q_Y+\alpha_Z q_Z)+\bigg(-(1.066~\textrm{km/s})\sin(t^\circ_{\rm lab})+(16.56~\textrm{km/s})\cos(t^\circ_{\rm lab})\bigg)\nonumber\\
&(\beta_X q_X+\beta_Y q_Y+\beta_Z q_Z)+\bigg(-\big[(1.066~\textrm{km/s})\cos(t^\circ_{\rm lab})+(16.56~\textrm{km/s})\sin(t^\circ_{\rm lab})\big]\cos(\lambda_{\rm lab})\nonumber\\
&+(7.077 ~\textrm{km/s})\sin(\lambda_{\rm lab})\bigg)(\gamma_X q_X+\gamma_Y q_Y+\gamma_Z q_Z).
\label{qdotSolar}
\end{align}
 Clearly, Eq.~\ref{qdotSolar} has a time dependence through $t^\circ_{\rm lab}$ and would be responsible of any daily modulation in the rate.

\subsection{Earth's revolution}

The velocity of the Earth's revolution around the sun is given in terms of the Sun ecliptic longitude $\lambda(t)$ as~\cite{Green}
\begin{align}
{\bf V}_{\rm {Earth Rev}}&=V_{\oplus}(\lambda(t)) [\cos\beta(x) \sin (\lambda(t)-\lambda_x) \hat{{\bf x}}_g \nonumber\\
&+ \cos\beta(y) \sin (\lambda(t)-\lambda_y) \hat{{\bf y}}_g + \cos\beta(z) \sin (\lambda(t)-\lambda_z) \hat{{\bf z}}_g],
\label{EarthRev}
\end{align}
where $V_{\oplus}=29.8$ km/s is the orbital speed of the Earth, $V_{\oplus}(\lambda(t))=V_{\oplus}[1-e \sin(\lambda(t)-\lambda_0)]$, $e=0.016722$, and $\lambda_0=13^\circ+1^\circ$ are the ellipticity of the Earth's orbit and the ecliptic longitude of the orbit's minor axis, respectively, and $\beta_i=(-5^\circ.5303, 59^\circ.575, 29^\circ.812)$ and $\lambda_i=(266^\circ.141, -13^\circ.3485, 179^\circ.3212)$ are the ecliptic latitudes and longitudes of the ($\hat{{\bf x}}_g$,$\hat{{\bf y}}_g$,$\hat{{\bf z}}_g$) axes, respectively.

The Sun's ecliptic longitude $\lambda(t)$ can be expressed as (p. 77  of Ref.~\cite{Lang} and Ref.~\cite{Green}),
\begin{equation}
\lambda(t)=L + (1^\circ .915 - 0^\circ.0048 T_0) \sin g+ 0^\circ .020 \sin 2g,
\end{equation}
where $L=281^\circ .0298 + 36000^\circ .77 T_0 + 0^\circ .04107 UT$ is the mean longitude of the Sun corrected for aberration, $g=357^\circ .9258 + 35999^\circ .05 T_0 + 0^\circ .04107 UT$ is the mean anomaly (polar angle of orbit).

Using Eq.~\ref{Equit-Gal}, we can transform from the galactic frame to the equatorial frame, and using Eq.~\ref{eq:labeq3} we can transform from the equatorial frame to the lab frame  ($\hat{\cal N},\hat{\cal W},\hat{\cal Z}$). Then we can use Eq.~\ref{Lab-Crystal} to write ${\bf V}_{\rm {Solar}}$ in terms of the crystal frame coordinates.

Thus, we can compute $\hat{\mathbf{q}} \cdot{\bf V}_{\rm {Earth Rev}}$ as
\begin{align}
\hat{\mathbf{q}} \cdot{\bf V}_{\rm {Earth Rev}}&=V_{\oplus}(\lambda(t)) \bigg\{\big[-\cos(t^\circ_{\rm lab})  \sin(\lambda_{\rm lab}) {\cal A}(t) -\sin(t^\circ_{\rm lab}) \sin(\lambda_{\rm lab}) {\cal B}(t) + \cos(\lambda_{\rm lab}) {\cal C}(t) \big]\nonumber\\
&\left(\alpha_X q_X+\alpha_Y q_Y+\alpha_Z q_Z\right) + \big[\sin(t^\circ_{\rm lab}) {\cal A}(t) -\cos(t^\circ_{\rm lab}) {\cal B}(t) \big]\left(\beta_X q_X+\beta_Y q_Y+\beta_Z q_Z\right)\nonumber\\
&+\big[\cos(t^\circ_{\rm lab})  \cos(\lambda_{\rm lab}) {\cal A}(t) + \sin(t^\circ_{\rm lab}) \cos(\lambda_{\rm lab}) {\cal B}(t) +\sin(\lambda_{\rm lab}) {\cal C}(t) \big] \left(\gamma_X q_X+\gamma_Y q_Y+\gamma_Z q_Z\right)\bigg\},
\label{qdotEarthRev}
\end{align}
where
\begin{align}
{\cal A}(t)&=(-0.06699) \cos\beta(x) \sin (\lambda(t)-\lambda_x) + (0.4927)\cos\beta(y) \sin (\lambda(t)-\lambda_y)\nonumber\\
&+ (-0.8676)\cos\beta(z) \sin (\lambda(t)-\lambda_z),\nonumber\\
{\cal B}(t)&= (-0.8728)\cos\beta(x) \sin (\lambda(t)-\lambda_x) + (-0.4503) \cos\beta(y) \sin (\lambda(t)-\lambda_y)\nonumber\\
&+ (-0.1883) \cos\beta(z) \sin (\lambda(t)-\lambda_z),\nonumber\\
{\cal C}(t)&= (-0.4835) \cos\beta(x) \sin (\lambda(t)-\lambda_x) + (0.7446) \cos\beta(y) \sin (\lambda(t)-\lambda_y)\nonumber\\
&+ (0.4602) \cos\beta(z) \sin (\lambda(t)-\lambda_z).
\end{align}

Eq.~\ref{qdotEarthRev} has a time dependence through $t^\circ_{\rm lab}$ and $\lambda(t)$ and would be responsible for any daily modulation in the rate.

\subsection{Earth's rotation}

Finally, we want to compute ${\bf V}_{\rm {Earth Rot}}$, the velocity of Earth's rotation around itself. We have
\begin{equation}
{\bf V}_{\rm {Earth Rot}}=-V_{\rm {RotEq}} \cos \lambda_{\rm lab} \hat{\cal W},
\end{equation}
where $V_{\rm {RotEq}}$ is the  Earth's rotation speed at the equator, and is defined as $V_{\rm {RotEq}}=2 \pi R_{\oplus}/({\rm {1~ sidereal~ day}})$. The Earth's equatorial radius is $R_{\oplus}=6378.137$ km, and one sidereal day is 23.9344696 hr$=86164$ s. therefore $V_{\rm {RotEq}}=0.465102$ km/s.

Using Eq.~\ref{Lab-Crystal} to write $ \hat{\cal W}$ in terms of the crystal frame coordinates, we can easily find $\hat{\mathbf{q}} \cdot {\bf V}_{\rm {Earth Rot}}$ as
\begin{equation}
\hat{\mathbf{q}} \cdot{\bf V}_{\rm {Earth Rot}}=-V_{\rm {RotEq}} \cos \lambda_{\rm lab} \left(\beta_X q_X+\beta_Y q_Y+\beta_Z q_Z\right).
\label{qdotEarthRot}
\end{equation}
There is no time dependence in Eq.~\ref{qdotEarthRot}, because it is written in the crystal frame, and both the lab and the crystal are rotating with the Earth.

\subsection{Total Velocity}
Now we can insert Eqs.~\ref{qdotGalacticRot},~\ref{qdotSolar},~\ref{qdotEarthRev} and ~\ref{qdotEarthRot} into Eq.~\ref{qdotV} to compute $\hat{\mathbf{q}} \cdot{\bf V}_{{\rm lab}}$. Inserting the values of $V_{\oplus}=29.8$ km/s, $\epsilon=23.439 ^\circ$ and $V_{\rm {RotEq}}=0.465$ km/s, we have (in km/s):
\begin{align}
\hat{\mathbf{q}} \cdot{\bf V}_{{\rm lab}}&=\bigg\{\bigg[-\cos(t^\circ_{\rm lab})~A(t)+\sin(t^\circ_{\rm lab})~B(t)\bigg]\sin\lambda_{\rm lab}
+C(t)~\cos\lambda_{\rm lab}\bigg\}\left(\alpha_X q_X+\alpha_Y q_Y+\alpha_Z q_Z\right)\nonumber\\
&+\bigg\{\sin(t^\circ_{\rm lab})~A(t)+\cos(t^\circ_{\rm lab})~B(t)-0.465 \cos \lambda_{\rm lab}\bigg\}\left(\beta_X q_X+\beta_Y q_Y+\beta_Z q_Z\right)\nonumber\\
&+\bigg\{\bigg[\cos(t^\circ_{\rm lab})~A(t)-\sin(t^\circ_{\rm lab})~B(t)\bigg]\cos\lambda_{\rm lab}+C(t)~\sin\lambda_{\rm lab}\bigg\}\left(\gamma_X q_X+\gamma_Y q_Y+\gamma_Z q_Z\right),
\label{qdotVlab}
\end{align}
where
\begin{align}
A(t)&=0.4927~ V_{\rm {Gal Rot}} - 1.066~\textrm{km/s} + (V_{\oplus}(\lambda(t)) {\cal A}(t),\nonumber\\
B(t)&=0.4503~ V_{\rm {Gal Rot}} + 16.56~\textrm{km/s} - (V_{\oplus}(\lambda(t)) {\cal B}(t),\nonumber\\
C(t)&=0.7445~ V_{\rm {Gal Rot}} + 7.077~\textrm{km/s} + (V_{\oplus}(\lambda(t)) {\cal C}(t).
\end{align}

\end{document}